\pdfoutput=1
\documentclass[useAMS,usenatbib]{mn2e}

\usepackage{amsmath,amssymb}
\usepackage{mathrsfs}
\usepackage{graphicx, color}
\usepackage{cprotect}
\def\mr#1{\mathrm{#1}}

\title[The growth of Abell 85]{The growth of the galaxy cluster Abell 85: mergers, shocks, stripping and seeding of clumping}
\author[Y. Ichinohe et al.]{Y. Ichinohe$^{1,2}$\thanks{E-mail: ichinohe@astro.isas.jaxa.jp}, N. Werner$^{3,4}$, A. Simionescu$^{1}$, S. W. Allen$^{3,4}$, R. E. A. Canning$^{3,4}$,
\newauthor
S. Ehlert$^{5}$, F. Mernier$^{6,7}$ and T. Takahashi$^{1,2}$\\
$^1$Institute of Space and Astronautical Science (ISAS),
JAXA, 3-1-1 Yoshinodai, Chuo, Sagamihara, Kanagawa, 252-5210 Japan\\
$^2$Department of Physics, Graduate School of Science,
University of Tokyo, 7-3-1 Hongo, Bunkyo, Tokyo, 113-0033 Japan\\
$^3$Kavli Institute for Particle Astrophysics and Cosmology, Stanford University,
452 Lomita Mall, Stanford, CA 94305-4085, USA\\
$^4$Department of Physics, Stanford University, 382 Via Pueblo Mall, Stanford, CA 94305-4060, USA\\
$^5$Kavli Institute for Astrophysics and Space Research, Massachusetts Institute of Technology,
77 Massachusetts Avenue, Cambridge, \\MA 02139, USA\\
$^6$SRON Netherlands Institute for Space Research, Sorbonnelaan 2, 3584, CA Utrecht, The Netherlands\\
$^7$Leiden Observatory, Leiden University, P.O. Box 9513, 2300 RA, Leiden, The Netherlands}
\begin{document}

\date{\today}

\pagerange{\pageref{firstpage}--\pageref{lastpage}} \pubyear{2014}

\maketitle

\label{firstpage}
\begin{abstract}
 We present the results of deep {\it Chandra}, {\it XMM-Newton} and {\it Suzaku} observations of the nearby galaxy cluster Abell~85, which is currently undergoing at least two mergers, and in addition shows evidence for gas sloshing which extends out to $r \approx 600$~kpc. One of the two infalling subclusters, to the south of the main cluster center, has a dense, X-ray bright cool core and a tail extending to the southeast. The northern edge of this tail is strikingly smooth and sharp (narrower than the Coulomb mean free path of the ambient gas) over a length of 200~kpc, while toward the southwest the boundary of the tail is blurred and bent, indicating a difference in the plasma transport properties between these two edges. The thermodynamic structure of the tail strongly supports an overall northwestward motion. We propose, that a sloshing-induced tangential, ambient, coherent gas flow is bending the tail eastward. The brightest galaxy of this subcluster is at the leading edge of the dense core, and is trailed by the tail of stripped gas, suggesting that the cool core of the subcluster has been almost completely destroyed by the time it reached its current radius of $r \approx 500$~kpc. The surface-brightness excess, likely associated with gas stripped from the infalling southern subcluster, extends toward the southeast out to at least $r_{500}$ of the main cluster, indicating that the stripping of infalling subclusters may seed gas inhomogeneities. The second merging subcluster appears to be a diffuse non-cool core system. Its merger is likely supersonic with a Mach number of $\approx1.4$.
\end{abstract}


\section{Introduction}
The growth of cosmic structure is driven by the gravity of dark matter, which dominates the mass in the Universe. Among gravitationally collapsed systems, clusters of galaxies are the largest and most recently formed (and still forming) objects, which evolve via accretion and successive mergers with smaller structures. During a merger, the infalling subsystem experiences ram pressure stripping and the stripped gas mixes with the ambient gas of the main cluster. Being heated to temperatures of 10$^7$-10$^8$~K via mixing and shock heating, the gas shines brightly at X-ray wavelengths. As most of the baryonic matter in clusters is in the form of this hot gas, X-ray observations are a powerful tool to investigate structure formation \citep{sarazin86,bohringer10,allen11}. However, the process by which stripped gas becomes virialized within the cluster is poorly understood and will depend on the microphysics of the intracluster medium (ICM). Recently, observations have shed new light on this process, indicating that the ICM in the cluster outskirts may often be clumpy, which suggests a slow or incomplete mixing \citep{simionescu11,urban11,morandi13,urban14}.

We observe various features associated with cluster mergers, with sizes ranging from $\sim$kpc to $\sim$Mpc. The three main types of features are shock fronts, cold fronts, and sloshing. When a subcluster moves supersonically through the ICM, it drives a shock front, which heats and compresses the gas, producing a sharp jump of thermodynamic properties (density, temperature, pressure and entropy) across the interface \citep{markevitch02, markevitch05, russell10}. When a cool dense gas parcel moves subsonically through the rarefied hot ambient gas, we see another type of a sharp contact discontinuity, a cold front, across which both the temperature and density exhibit a jump, but with a contrary relation to the shock front, resulting in an almost continuous pressure profile \citep{markevitch00, vikhlinin01a, mazzotta01, sun02}. Thirdly, when a merger occurs with a non-zero impact parameter, gas sloshing can be triggered, and will manifest itself by a characteristic morphology (a spiral pattern or concentric arcs, depending on the line-of-sight orientation with respect to the merger plane, \citealt{tittley05,ascasibar06,roediger11}), with cold fronts delineating its edge \citep{vikhlinin01b}. Cold fronts due to ICM sloshing are present in many cooling-core clusters \citep{ghizzardi10}. These features are useful diagnostic tools for studying gas dynamics during the clusters' hierarchical evolution \citep{markevitch07}.

Abell~85 ($z=0.055$, \citealt{oegerle01}) is one of the brightest galaxy clusters in the X-ray sky \citep{edge90}. It has been observed with various X-ray instruments, e.g. {\it Chandra} \citep{kempner02}, {\it XMM-Newton} \citep{durret03,durret05b}, {\it Suzaku} \citep{tanaka10}, and also in other wavelengths, e.g. radio \citep{slee01,schenck14}. The main cluster hosts an X-ray bright, metal rich, ``cool core'', and a recent optical study \citep{lopez-cruz14} has shown its brightest central galaxy, Holm 15A, is the largest one known so far.

Abell~85 is also one of the most complex interacting systems known. Currently, at least two subclusters are falling into the main cluster: one from the south and the other from the southwest \citep{kempner02,durret05b}. Additionally, the ICM distribution shows evidence for sloshing \citep{lagana10}. This remarkable dynamical activity makes Abell~85 an excellent system to investigate the process of cluster growth.

The subcluster falling in from the south (S subcluster) also has a bright central galaxy surrounded by cool X-ray gas. It has a clear tail structure, extending to the southeast, previously seen in both {\it Chandra} and {\it XMM-Newton} images. Previous {\it Suzaku} observations \citep{tanaka10} detected a possible shock front in the northeast of the subcluster.

The second subcluster located towards the southwest (SW subcluster) has not previously been studied in detail, but \citet{schenck14} revealed the gas in the interface between the main cluster and this subcluster to have a high temperature. This region may be caused by a shock front induced by the SW subcluster merger, but associated radio emission has not been detected except for relic structures which are located at $\sim$150~kpc west of the interface region.

The outline of this paper is as follows: In Section \ref{observations and data reduction}, we describe our new, deep {\it Chandra} observations, archival {\it XMM-Newton} data, and new {\it Suzaku} observations that probe the cluster out to $r_{200}$ along the infall direction of the S subcluster. We also describe here the data reduction including the imaging analysis, and the production of the thermodynamic maps and deprojected profiles. We describe the results in Section \ref{results}, and discuss them in Section \ref{discussion}. Conclusions are presented in Section \ref{conclusions}. We assume $H_0 = 70\ \mr{km\ s^{-1}~Mpc^{-1}}$, $\Omega_M = 0.3$ and $\Omega_\Lambda = 0.7$ ($1^{\prime\prime} = 1.069~\mr{kpc}$ at $z = 0.055$) throughout this paper. Errors are given at the 1$\sigma$ level unless otherwise stated.

\section[]{Observations and data reduction}\label{observations and data reduction}
\begin{table}
 \centering
 \caption{Summary of the observations. Net exposure times are the values after cleaning.}
\vspace{0.3cm}
\resizebox{\hsize}{!}{
 \begin{tabular}{cccc}
  \hline
  Satellite & ObsID & Date & Net exposure time (ks)\\
  \hline
  {\it Chandra} & 15173 & 2013-08-14 & 39\\
  {\it Chandra} & 15174 & 2013-08-09 & 36\\
  {\it Chandra} & 16263 & 2013-08-10 & 37\\
  {\it Chandra} & 16264 & 2013-08-17 & 34\\
  {\it Chandra} &   904 & 2000-08-19 & 37\\
  {\it XMM-Newton} & 0065140101 & 2002-01-07 & 12\\
  {\it XMM-Newton} & 0065140201 & 2002-01-07 & 12\\
  {\it XMM-Newton} & 0723802101 & 2013-06-16 & 97\\
  {\it XMM-Newton} & 0723802201 & 2013-06-18 & 98\\
  {\it Suzaku} & 801041010 & 2007-01-05 & 82\\
  {\it Suzaku} & 807135010 & 2012-12-31 & 127\\
  {\it Suzaku} & 807136010 & 2013-01-03 & 50\\
  \hline
 \end{tabular}}
 \label{observations}
\end{table}

\subsection{{\it Chandra} observations}
The data from four newly observed pointings (ObsID 15173, 15174, 16263, 16264) and one archival observation (ObsID 904) were processed in the standard manner using the {\small CIAO} software package (version 4.6) and the {\small CALDB} version 4.6.2 (see e.g. \citet{million10} for details). The net exposure times after removing bad pixels and deflaring the data are summarized in Table \ref{observations}. Blank-sky background files were processed in a similar manner and were scaled by the ratio of photon counts in the data to those of the background in the energy range 9.5-12~keV.

\subsubsection{Imaging analysis}
\begin{figure*}
 \begin{minipage}{0.495\hsize}
  \begin{center}
   \includegraphics[width=3.35in]{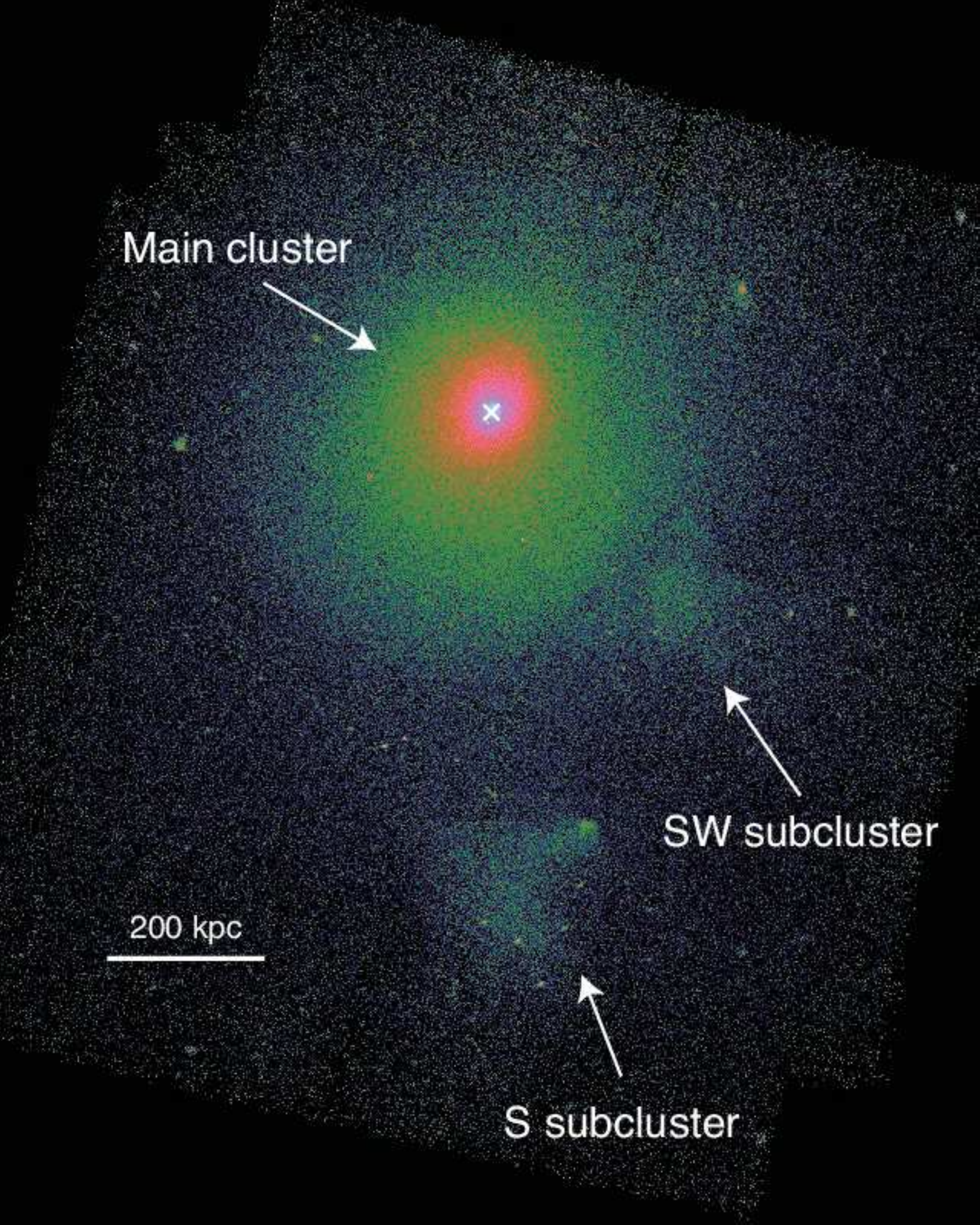}
  \end{center}
 \end{minipage}
 \begin{minipage}{0.495\hsize}
  \begin{center}
   \includegraphics[width=3.35in]{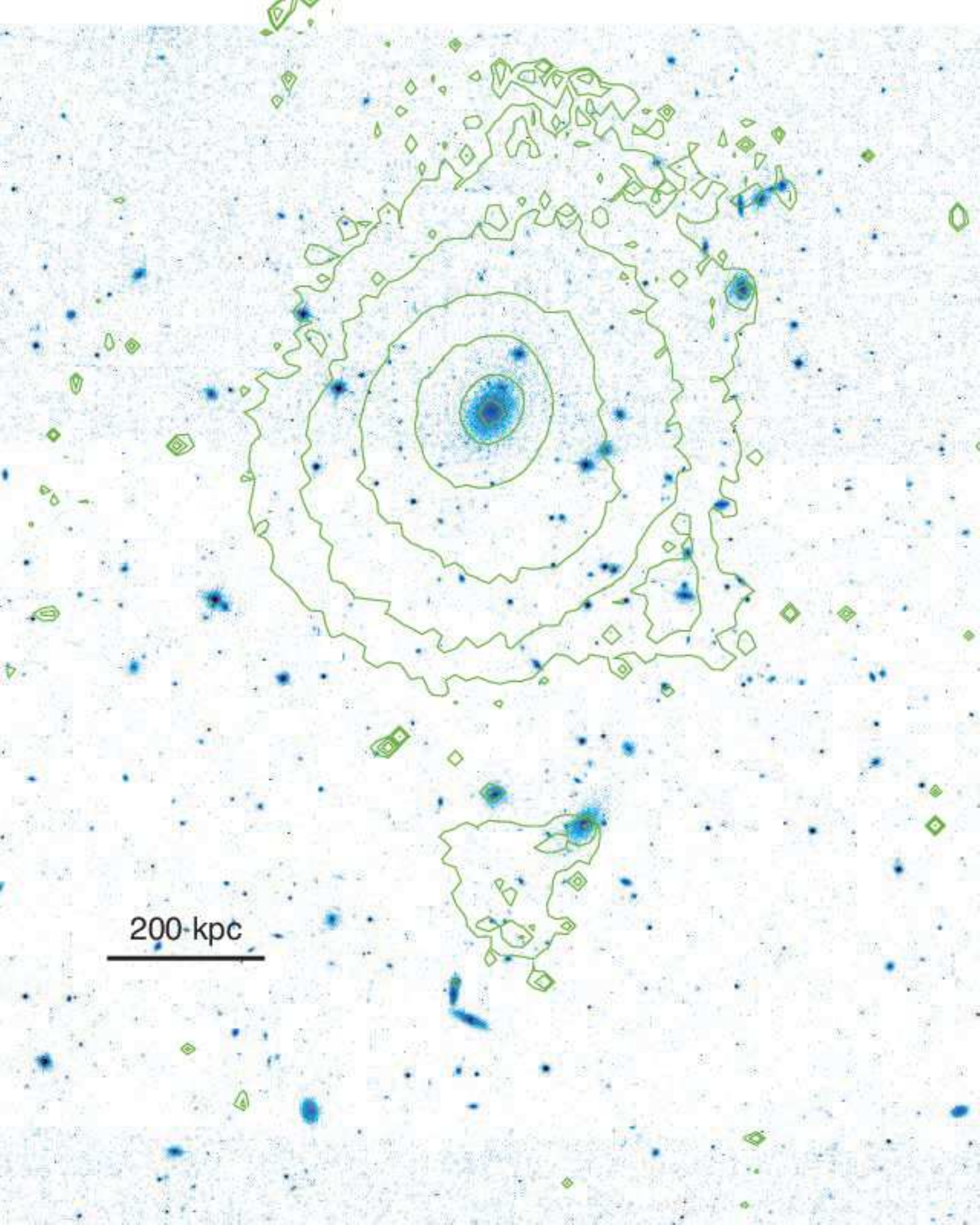}
  \end{center}
 \end{minipage}
 \caption{{\it Left}: $\sigma=0.98~\mr{arcsec}$ Gaussian smoothed, exposure and vignetting corrected, background subtracted {\it Chandra} image of Abell~85 (0.6-7.5~keV). The cross represents the position of the central cD galaxy of the main cluster. {\it Right}: SDSS r-band optical image of the same sky region overlaid with {\it Chandra} X-ray contours.}
 \label{img_chandra}\label{img_sdss}
\end{figure*}
\begin{figure*}
 \begin{minipage}{0.495\hsize}
  \begin{center}
   \includegraphics[width=3.35in]{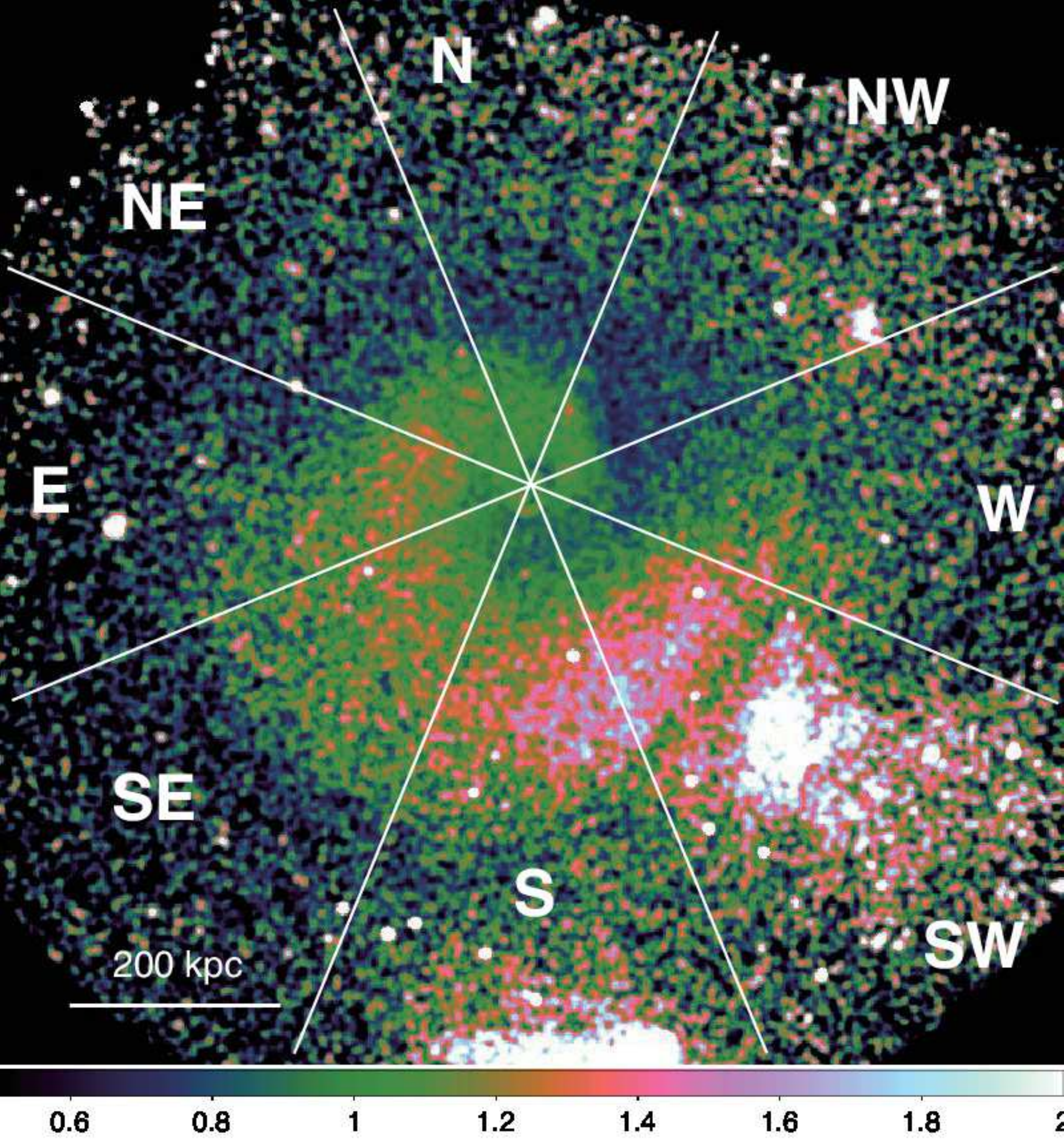}
  \end{center}
 \end{minipage}
 \begin{minipage}{0.495\hsize}
  \begin{center}
   \includegraphics[width=3.35in]{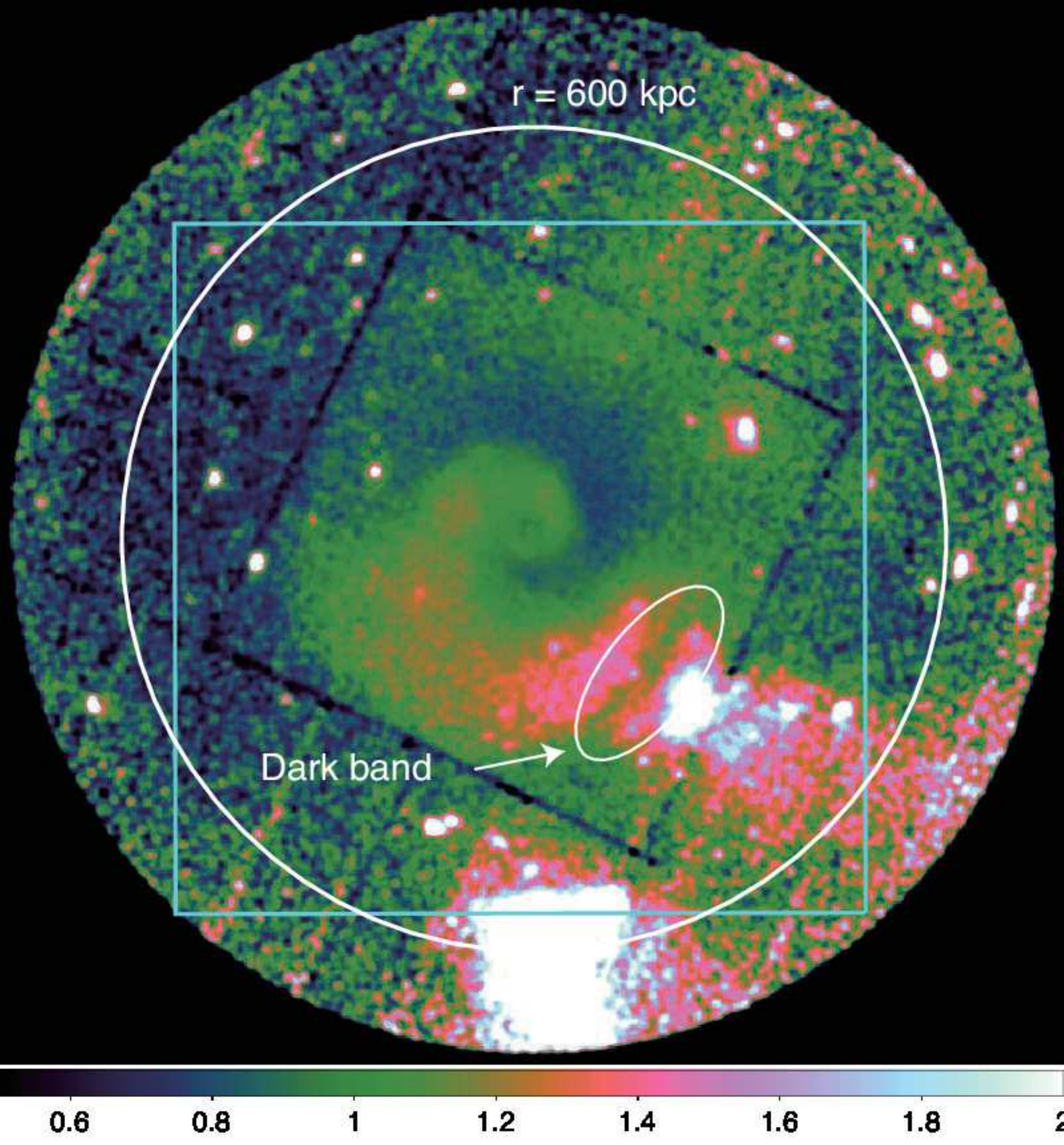}
  \end{center}
 \end{minipage}
 \caption{{\it Left}: $\sigma=5.6~\mr{arcsec}$ Gaussian smoothed, {\it Chandra} relative deviation image with respect to the best-fitting 2D elliptical double beta model.  We illustrate the eight directions for which projected radial profiles are extracted in Section \ref{sloshing} with the white lines. {\it Right}: $\sigma=8.3~\mr{arcsec}$ Gaussian smoothed, {\it XMM-Newton} relative deviation image. The white circle denotes the radius of 600~kpc from the main cluster core, and the cyan square denotes the FOV of the {\it Chandra} relative deviation image. There is a dark narrow structure (Dark band) at the interface region of the main cluster and the SW subcluster.}
 \label{residual_image}
\end{figure*}
Fig. \ref{img_chandra} shows the exposure and vignetting corrected, background subtracted {\it Chandra} image (left, 0.6-7.5~keV), created by combining background subtracted images and corresponding exposure maps for 13 narrow energy bands (0.6-0.8, 0.8-1.0, 1.0-1.2, 1.2-1.5, 1.5-2.0, 2.0-2.5, 2.5-2.75, 2.75-3.0, 3.0-3.5, 3.5-4.0. 4.0-6.0, 6.0-7.0, 7.0-7.5, all in keV). Also shown is the SDSS optical r-band image of the same field overlaid by X-ray contours (right).

To emphasize the small azimuthal asymmetries which cannot be clearly seen in the original images because of the overall surface brightness gradient, we created relative deviation images of the system with respect to the best-fitting 2D elliptical double beta model. We used {\small SHERPA} for fitting the sum of two 2D elliptical beta models: \verb-beta2d+beta2d-. The centre positions (\verb+xpos+, \verb+ypos+), ellipticities (\verb+ellip+) and the roll angles (\verb+theta+) were bound between the two beta models. Based on the {\it Chandra} image, the best fit ellipticity of the cluster is 0.17, and its long axis extends at 25 degree clockwise from the north. After best-fitting parameters were determined, we divided the original images by the best-fitting models. In Fig. \ref{residual_image}, we show the relative deviation images.

\subsubsection{Thermodynamic mapping}\label{thermodynamic_mapping}
\begin{figure*}
 \begin{minipage}{0.49\hsize}
  \includegraphics[width=3.25in]{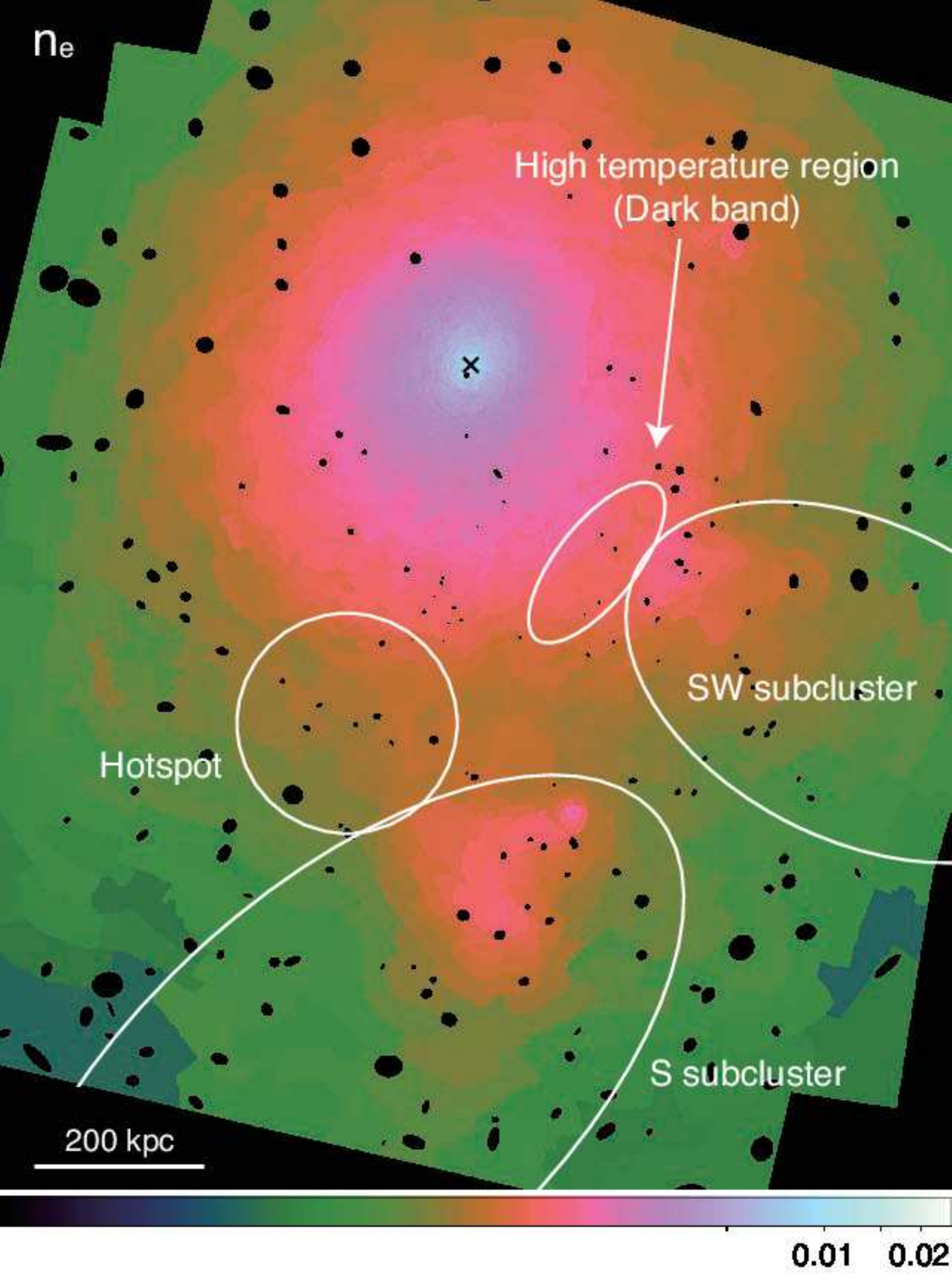}
 \end{minipage}
 \begin{minipage}{0.49\hsize}
  \includegraphics[width=3.25in]{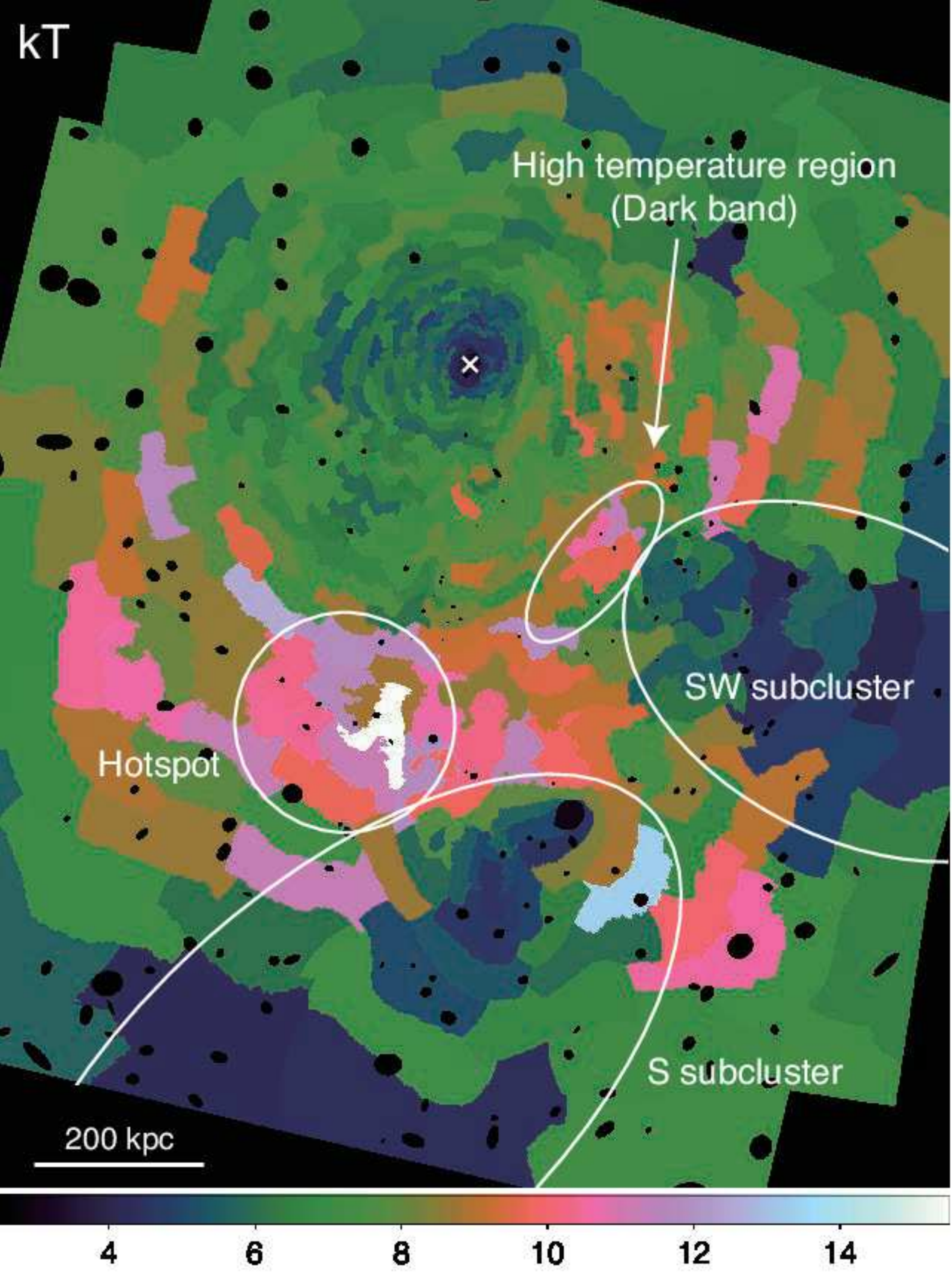}
 \end{minipage}
 \begin{minipage}{0.49\hsize}
  \includegraphics[width=3.25in]{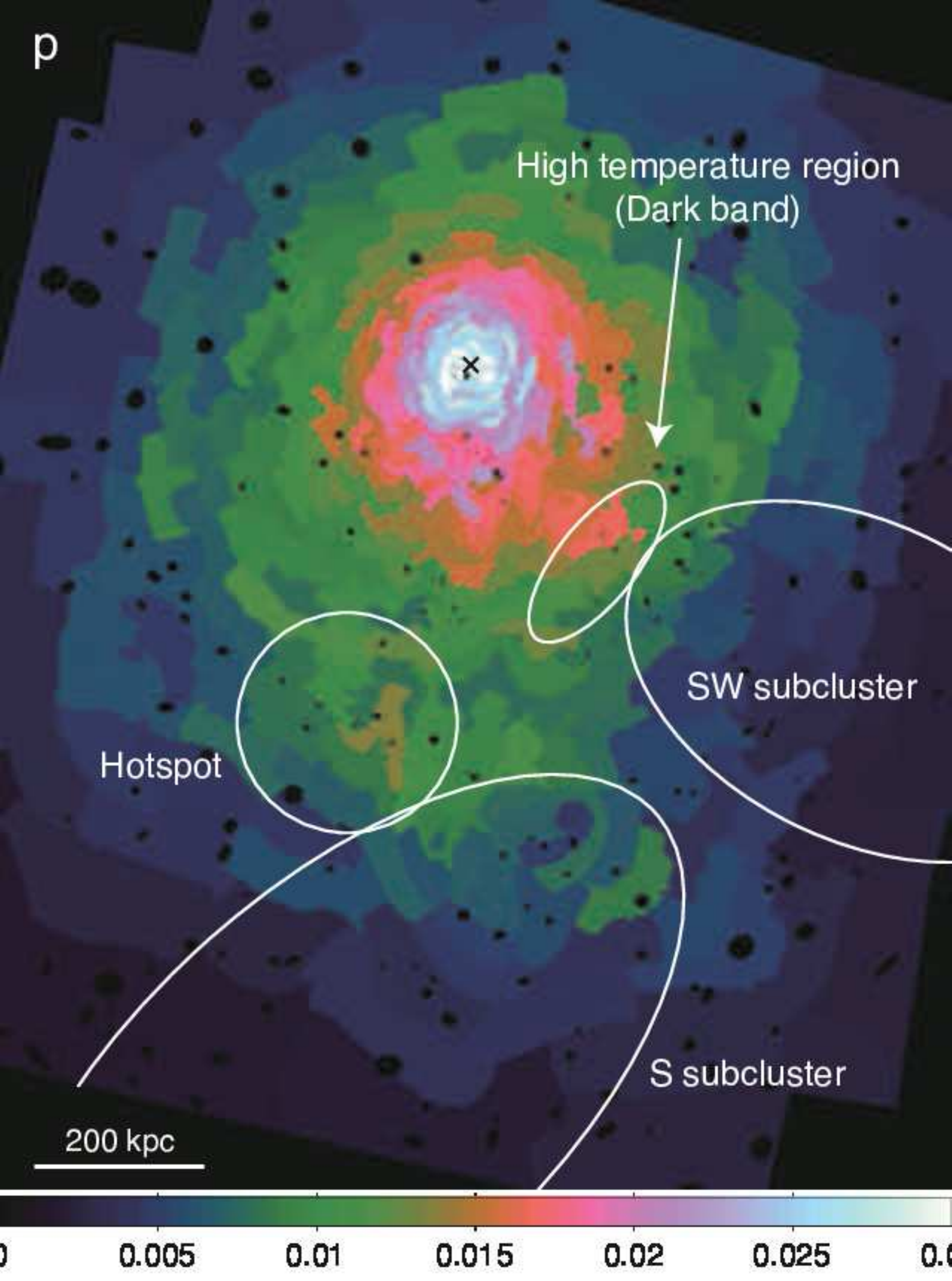}
 \end{minipage}
 \begin{minipage}{0.49\hsize}
  \includegraphics[width=3.25in]{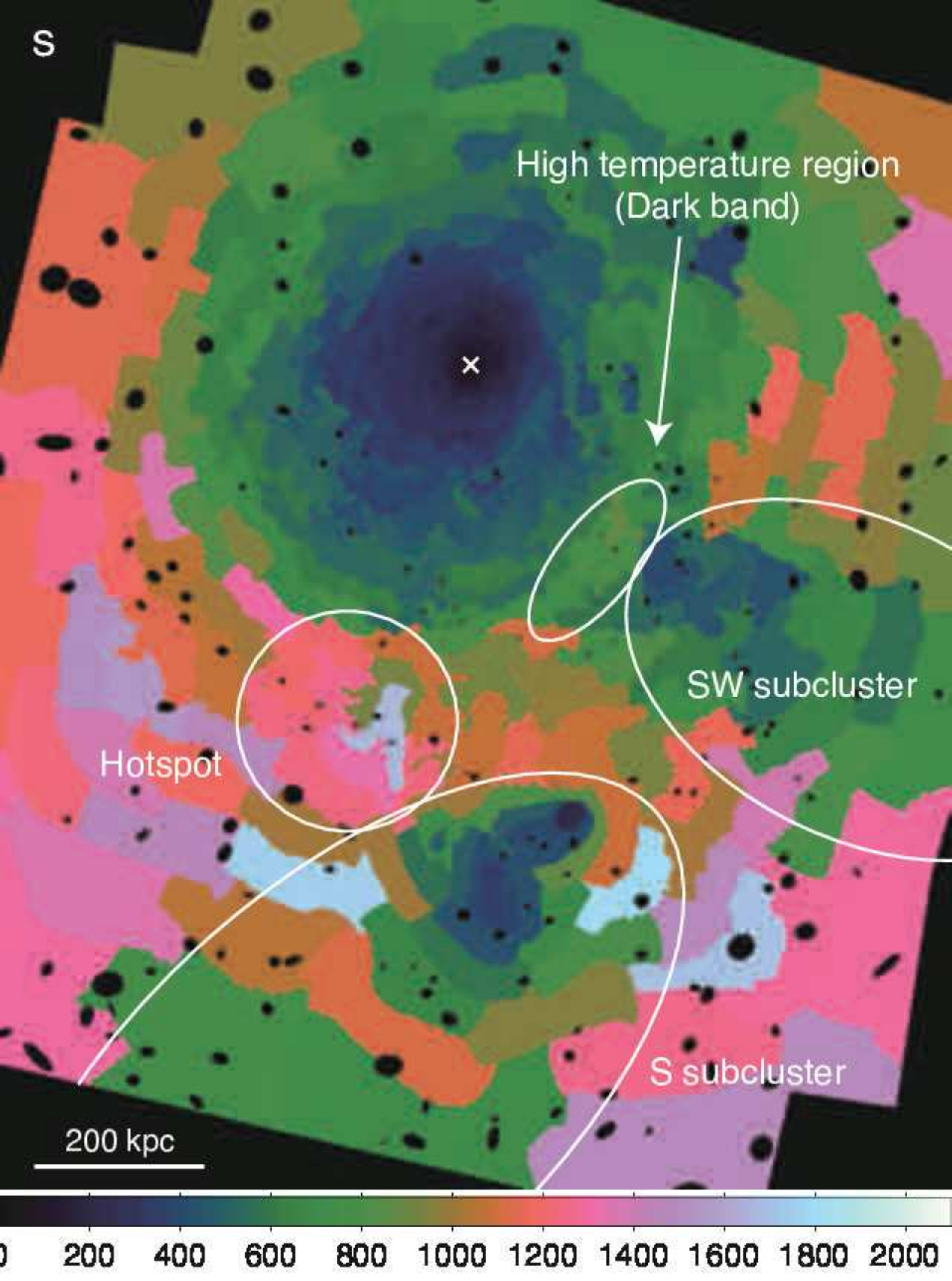}
 \end{minipage}
 \caption{Projected thermodynamic maps of Abell~85. The S/N values used to create these images were 33 (around 1100 counts) for the density map and 70 (around 4900 counts) for others. Typical errors were 5~per~cent for density and 10~per~cent for other quantities. We assume a uniform line-of-sight depth of $l=1$~Mpc over the entire field of view. The unit of density (left top), temperature (right top), pressure (bottom left) and entropy (bottom right) are $\mr{cm}^{-3}\times(l/1~\mr{Mpc})^{-1/2}$, keV, $\mr{keV cm}^{-3}\times(l/1~\mr{Mpc})^{-1/2}$ and $\mr{keV cm}^2\times(l/1~\mr{Mpc})^{1/3}$, respectively. The cross denotes the central cD galaxy of the main cluster. Two subclusters (S subcluster and SW subcluster) and their tails are clealry seen. High-temperature regions (Hotspot and Dark band) are also observed.}
 \label{map}
\end{figure*}

\begin{figure*}
 \begin{minipage}{0.49\hsize}
  \includegraphics[width=3.25in]{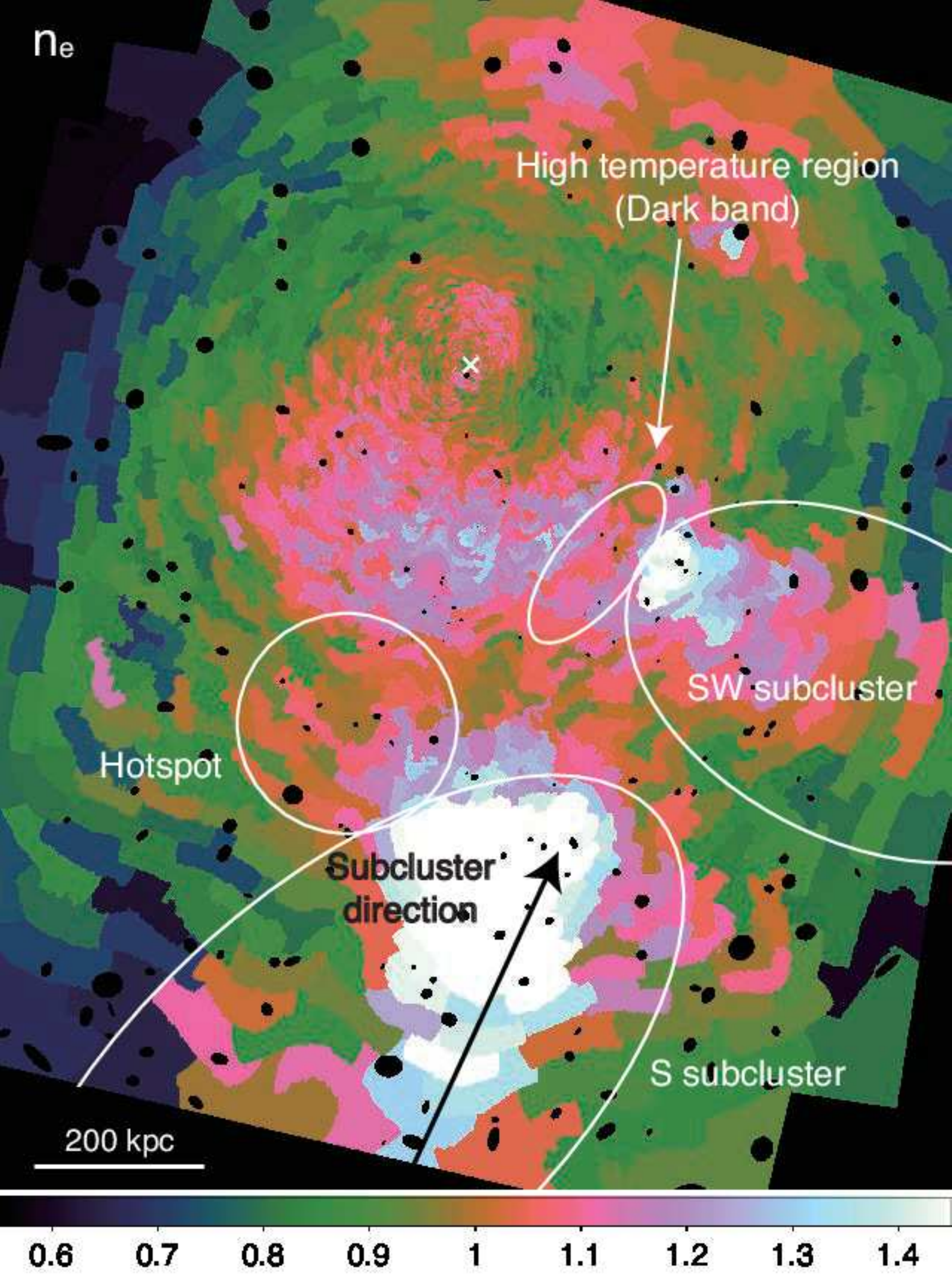}
 \end{minipage}
 \begin{minipage}{0.49\hsize}
  \includegraphics[width=3.25in]{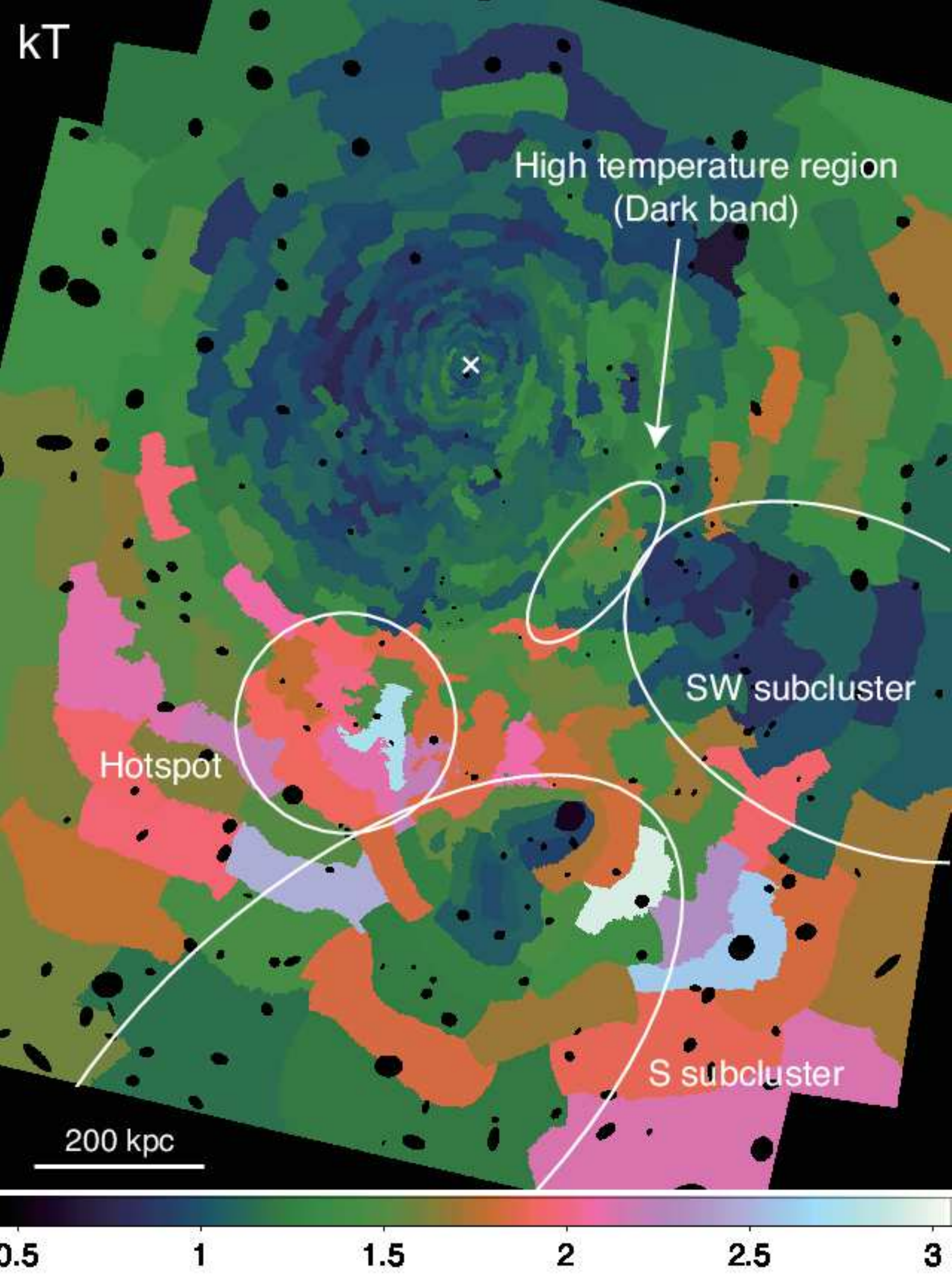}
 \end{minipage}
 \begin{minipage}{0.49\hsize}
  \includegraphics[width=3.25in]{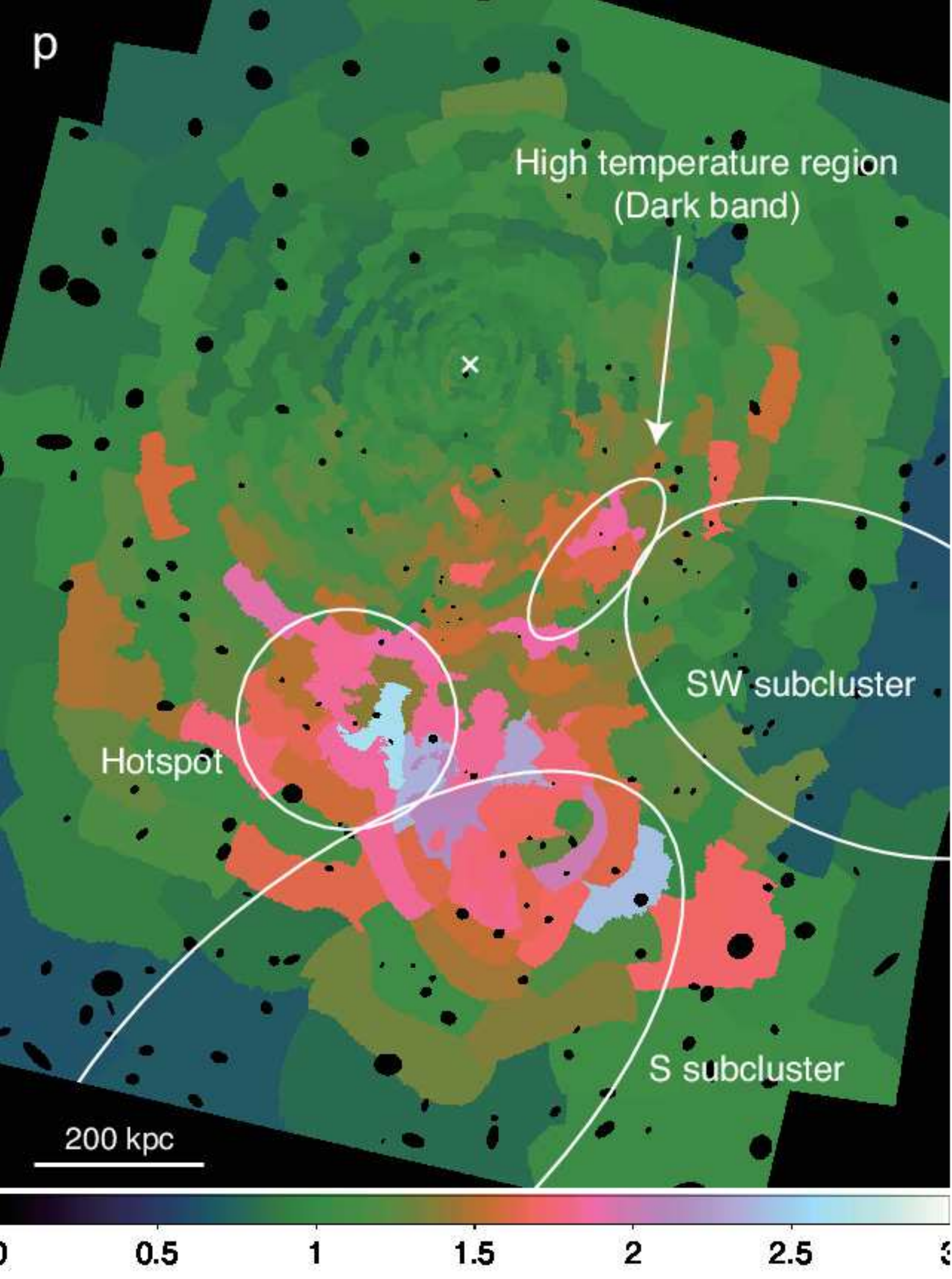}
 \end{minipage}
 \begin{minipage}{0.49\hsize}
  \includegraphics[width=3.25in]{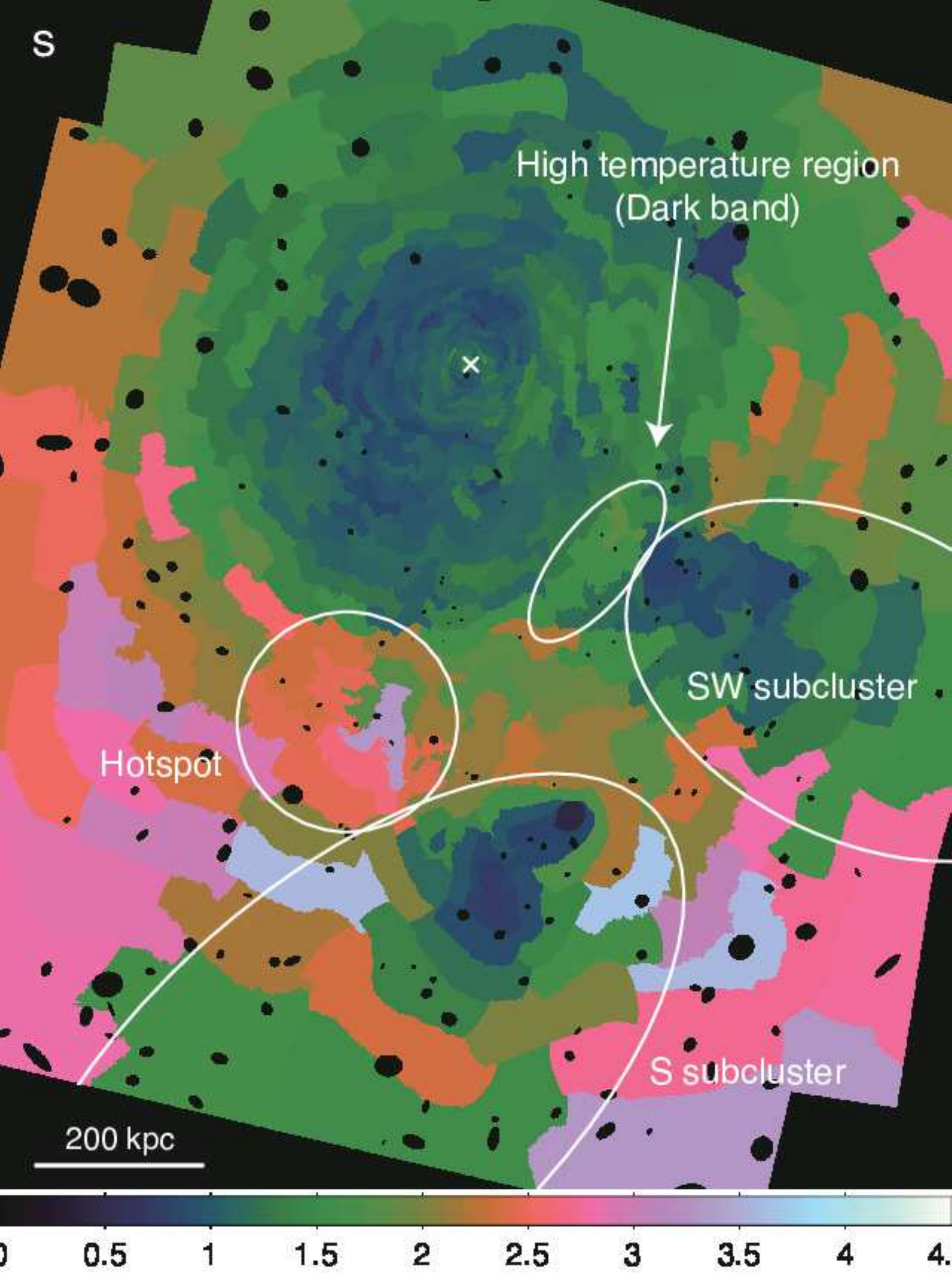}
 \end{minipage}
 \caption{Trend-divided thermodynamic maps of Abell~85. The arrangement is same as Fig. \ref{map}.}
 \label{map_tdiv}
\end{figure*}

We conducted thermodynamic mapping to explore the properties of the ICM. The thermodynamic maps are shown in Fig. \ref{map}. When creating these maps, we adopted the contour binning algorithm \citep{sanders06} which divides the field of view into smaller regions, each of which has similar signal-to-noise (S/N) ratio. Point sources were detected and excluded using the \verb+wavdetect+ tool with the scales of 1, 2, 4, 8, 16 pixels. We visually eliminated the false detections within 200~arcsec from the main cluster core and the S subcluster core.

For each of the small regions, we fitted the spectra using SPEX (version 2.04.01) \citep{kaastra96}. We used the model of a redshifted (\verb+reds+) collisional ionisation equilibrium plasma (\verb+cie+) absorbed by Galactic absorption (\verb+hot+). We used the chemical abundance table determined by \citet{grevesse98}. The redshift was fixed to 0.055 and the hydrogen column density was set to $2.8\times 10^{20}$~cm$^{-2}$, determined by the LAB (Leiden/Argentine/Bonn) radio HI survey \citep{kalberla05}. The temperature $kT$ was derived directly from the fitting results, the electron density $n_{\rm e}$ was calculated from the normalization of the \verb+cie+ model assuming the line-of-sight depth of the ICM is constant over the field of view (with a value of 1~Mpc) and the electron density $n_{\rm e}$ and ion density $n_X$ satisfy a relation $n_{\rm e} = 1.2 n_X$. The gas pressure $p$ and the entropy $s$ were calculated as $p = n_{\rm e}kT$ and $s = kTn_{\rm e}^{-2/3}$ respectively. Note that, although the uniform line-of-sight depth assumption is arbitrary, it should not significantly bias conclusions based on azimuthal variations in the thermodynamic maps. On the other hand, because we observe emission measure weighted thermodynamic quantities, projection effects in the presence of multiphase gas may bias our measurements.

The S/N ratios chosen to determine the region sizes were 33 (around 1100 counts) for the density and 70 (around 4900 counts) for the temperature and other secondary quantities. We obtain typical errors of 5~per~cent for the density map, and 10~per~cent for the temperature, pressure and entropy maps.

In order to emphasize the small azimuthal variations of these quantities, we also created the trend-divided thermodynamic maps shown in Fig. \ref{map_tdiv}. The scatter plots of each physical quantity versus distance from the main cluster core were fitted with a function using the form $f(r)=A(1+(r/B)^2)^{(-3C/2)}(1+(r/D)^2)^{(-3E/2)}$, where $A$, $B$, $C$, $D$, $E$ are free parameters, and $r$ is the distance from the cluster centre. After the $f(r)$ was determined, the typical values of the respective physical quantities were calculated for every region taking $r$ as the distance between the cluster centre and the geometrical centre of the region. Trend-divided maps were then created by dividing the original quantities by the typical value at the given radius.

\subsection{{\it XMM-Newton} observations}

Three out of the four Abell 85 observations with \textit{XMM-Newton} \citep{jansen01} were pointed at the cluster center and one observation was aimed  $\sim$25~arcmin southeast of the core (Table \ref{observations}), allowing us to look for extended emission in the south. The two deepest exposures were taken for the CHEERS\footnote{CHemical Evolution Rgs cluster Sample} project (de Plaa \& Cheers Collaboration in prep.); the two older exposures were previously analysed by \citet{durret03}. In all the observations the EPIC instruments were operating in Full Frame mode, except in the two deepest exposures where pn was operating in Extended Full Frame mode.

\begin{figure*}
 \begin{center}
  \includegraphics[width=4.3in]{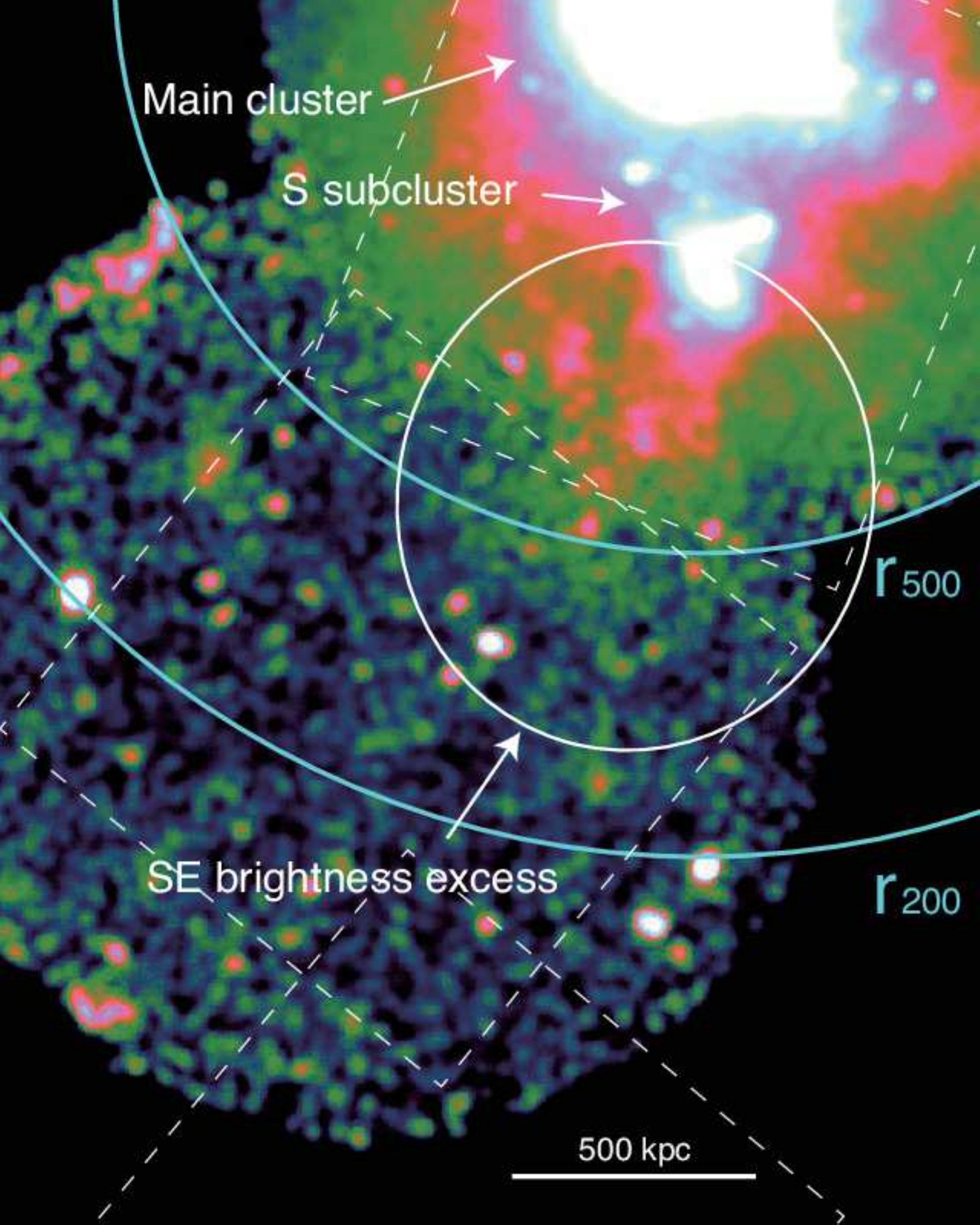}
  \caption{$\sigma=25~\mr{arcsec}$ Gaussian smoothed, exposure map corrected {\it XMM-Newton} EPIC (MOS+pn) image of Abell 85. The four \textit{XMM-Newton} datasets have been combined (Table \ref{observations}). The overlaid white dashed squares represent the {\it Suzaku} pointings.}
  \label{img_xmm}
 \end{center}
\end{figure*}

The data were reduced using the \textit{XMM-Newton} Science Analysis System ({\small SAS}) v13. The data reduction is adapted from Mernier et al. (submitted). In summary, after having processed the data using the {\small SAS} tasks \texttt{emproc} and \texttt{epproc}, we filtered the event files to exclude the soft-proton flares by building 2-$\sigma$ clipping Good Time Interval (GTI) files, both in the hard band (10-12~keV for MOS and 12-14~keV for pn) and the broad 0.3-10~keV band, because flares also occur in the soft band \citep{deluca04}. We then created one image per instrument and per observation. The image extraction was restricted to the 400--1250 eV band, where the instrumental background is relatively low and uniform across the detectors. Using closed-filter observations, rescaled using the out-of-field-of-view count rates, we corrected each image for instrumental background. We combined the resulting images and corrected them for exposure and vignetting. The final image is shown in Fig. \ref{img_xmm}. 

The relative deviation image shown in the right panel of Fig.~\ref{residual_image} has been produced by dividing the final flat-fielded {\it XMM-Newton} image with the best-fit 2D elliptical double beta model using the {\small SHERPA} package. The best fit parameters of the model are consistent with the values obtained by fitting the {\it Chandra} image.

\subsection{{\it Suzaku} observations}
We reduced the data from three {\it Suzaku} observations of Abell 85: one pointing centred on the core of the cluster, and two additional pointings offset approximately 26 and 43 arcmin south of the cluster's X-ray peak. The data are summarized in Table \ref{observations} (see also Fig. \ref{img_xmm} for the pointing locations). Here, we concentrate on the data obtained with the X-ray Imaging Spectrometer (XIS) detectors 0, 1, and 3.

The data were reduced using the tools available in the {\small HEAsoft} package (version 6.16) to create a set of cleaned event lists with hot or flickering pixels removed. All standard recommended screening criteria were applied\footnote{Arida M., XIS Data Analysis, http://heasarc.gsfc.nasa.gov/docs/suzaku/analysis/abc/node9.html (2010).}. We only included observation periods with the geomagnetic cut-off rigidity COR$>$6~GV. We used the latest calibration files that account for the modified non-X-ray background of the XIS1 detector following the increase in the charge injection level of 2011 June 1; in addition, for the XIS1 spectral analysis, we excluded two columns adjacent on each side of the charge-injected columns (the standard is to exclude one column on either side). We followed this methodology because the injected charge may leak into these additional columns and cause an increase in the instrumental background. We applied the latest contamination layer calibration from 2013 August 13.

We used the {\it Suzaku} mosaic to extract spectra from annuli centred on the cluster centre, $(\alpha,\delta)=(0:41:50.756,-9:18:03.49)$. Bright point sources were identified visually, cross-checked using the existing {\it XMM-Newton} mosaic, and removed from the analysis. The S subcluster seen in the {\it Chandra} image was removed as well. The projected and deprojected profiles of the thermodynamic properties were obtained with the {\small XSPEC} spectral fitting package \citep[version 12.8.2;][]{arnaud96}, employing the modified C-statistic estimator. We used the \texttt{projct} model to deproject the data under the assumption of spherical symmetry. Sets of spectra extracted from concentric annuli were modelled simultaneously in the 0.7-7.0~keV band. We modelled each shell as a single-temperature thermal plasma in collisional ionization equilibrium using the \verb+apec+ code \citep{smith01}, with the temperature and spectrum normalization as free parameters. The {\it Suzaku} spectrum normalizations throughout this paper are normalised to an extraction area of $20^2\pi$ arcmin$^2$. Unless otherwise noted, the metallicity was set to 0.3 Solar \citep[see][]{werner13}. The Galactic absorption column density was fixed to the average value measured from the LAB survey \citep{kalberla05}.

The instrumental background was subtracted in the standard way, using the task \texttt{xisnxbgen} which constructs a background spectrum based on the latest Night Earth observation files. The cosmic X-ray background (CXB) model consisted of a power-law emission component that accounts for the unresolved population of point sources, one absorbed thermal plasma model for the Galactic halo (GH) emission, and an unabsorbed thermal plasma model for the Local Hot Bubble (LHB). The parameters of the power-law were fixed based on the best-fit results obtained by \citet{simionescu13}, who analysed regions free of cluster emission in a large {\it Suzaku} mosaic centred around the Coma Cluster. We note that similar values for the CXB power-law are also obtained by \citet{urban14} who analysed a {\it Suzaku} Key project mosaic centred on the Perseus Cluster. The parameters for the LHB were fixed based on the all-sky average reported by \citet{kuntz00} using {\it ROSAT} data. Since the temperature of this component is only 0.08~keV, the signal that it contributes in the energy band 0.7-7.0~keV used for spectral fitting is negligible, therefore the best-fitting results are practically insensitive to large uncertainties on the normalization of this component. The temperature and normalization of the GH component were determined by fitting {\it Suzaku} spectra from an annulus free of cluster emission with inner and outer radii of 45 and 52~arcmin from the cluster core. The metallicity of the GH and LHB components were both assumed to be Solar.

The pointing in our mosaic located furthest from the centre of Abell 85 (the 46~arcmin offset), which was used to determine the best-fitting GH parameters, contains an X-ray bright star within the field of view that could be affecting the results. To mitigate this, we extracted a spectrum of this star and empirically modelled it with a power-law plus \verb+apec+ model; we then ran ray-tracing simulations to estimate the level and spectrum of the scattered light from this source, and included this in our spectral fitting of the GH parameters. We find that the contribution of the stray light from the star amounts to only 5 per cent of the total cosmic X-ray background flux in the 0.7-2.0~keV band. However, adding a cluster component above the CXB in the outer pointing in the mosaic in an annulus spanning 35-39~arcmin, we find that the contribution from the star can represent up to 30 per cent of this additional signal. We thus do not report any measurements of the cluster thermodynamic properties based on the outermost pointing -- only the centre and the 23~arcmin offset observations are used to derive the cluster properties.

\section{Results}\label{results}
In the {\it Chandra} image and the thermodynamic maps (Fig. \ref{img_chandra} left, Fig. \ref{map}, Fig. \ref{map_tdiv}), two subclusters are clearly seen in addition to the brightest main cluster \citep[see also e.g.][]{durret05b}; one is $\sim$500~kpc south of the main cluster core (S subcluster) and the other is $\sim$350~kpc southwest of the core (SW subcluster).

\subsection{The southern merger}\label{ssub}
\subsubsection{S subcluster and a ``hotspot''}
\begin{figure*}
 \begin{minipage}{0.33\hsize}
  \begin{center}
   \includegraphics[width=2.3in]{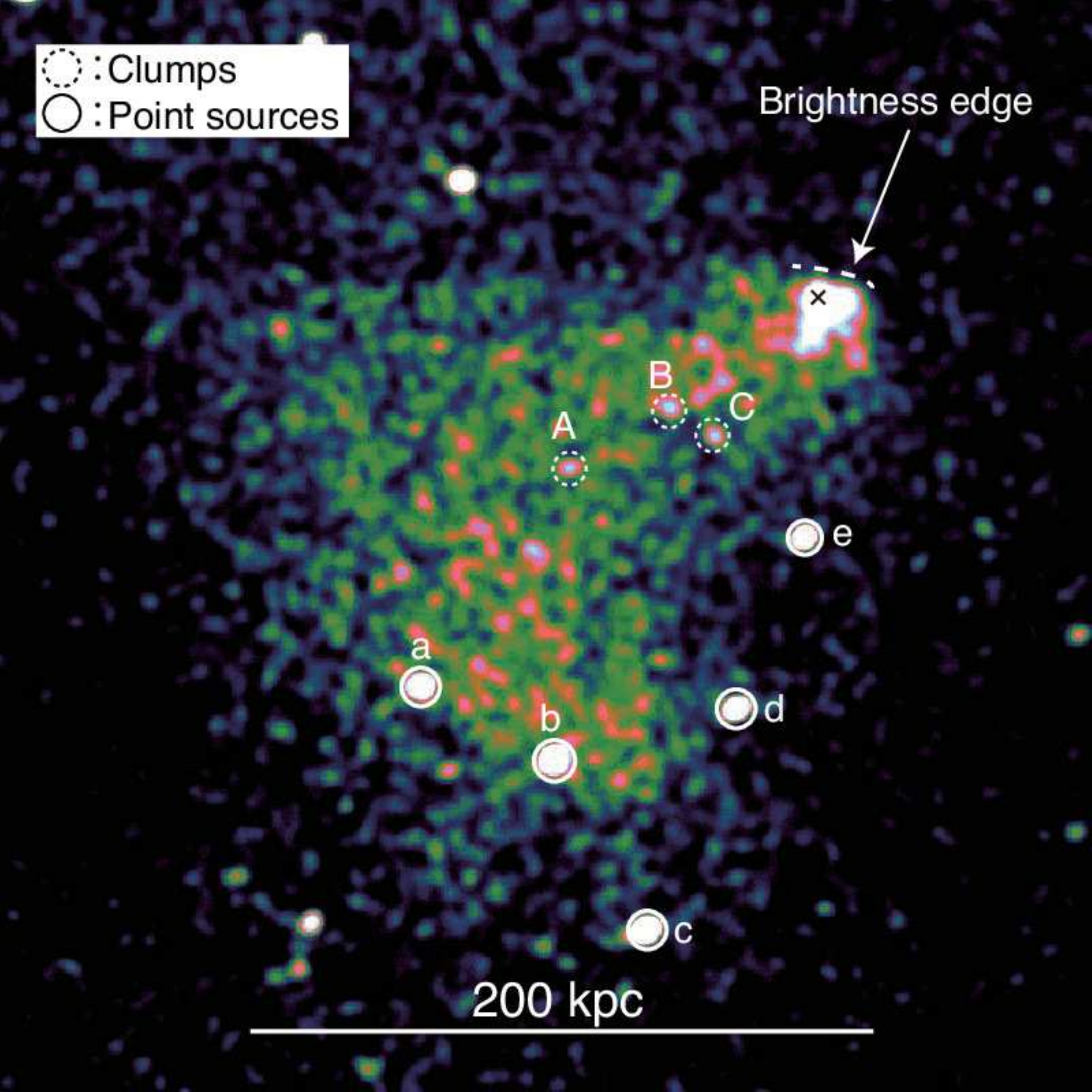}
  \end{center}
 \end{minipage}
 \begin{minipage}{0.33\hsize}
  \begin{center}
   \includegraphics[width=2.3in]{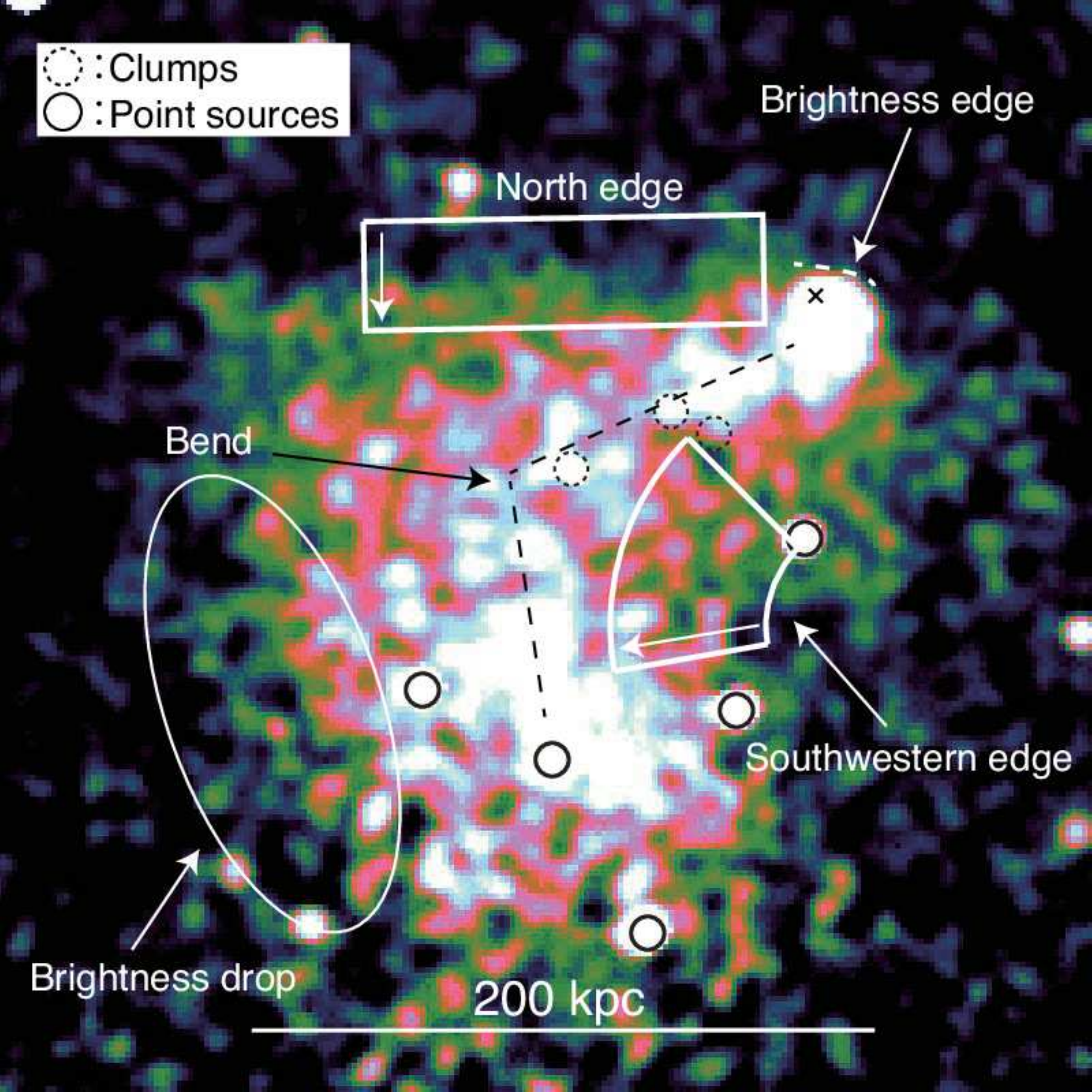}
  \end{center}
 \end{minipage}
 \begin{minipage}{0.33\hsize}
  \begin{center}
   \includegraphics[width=2.3in]{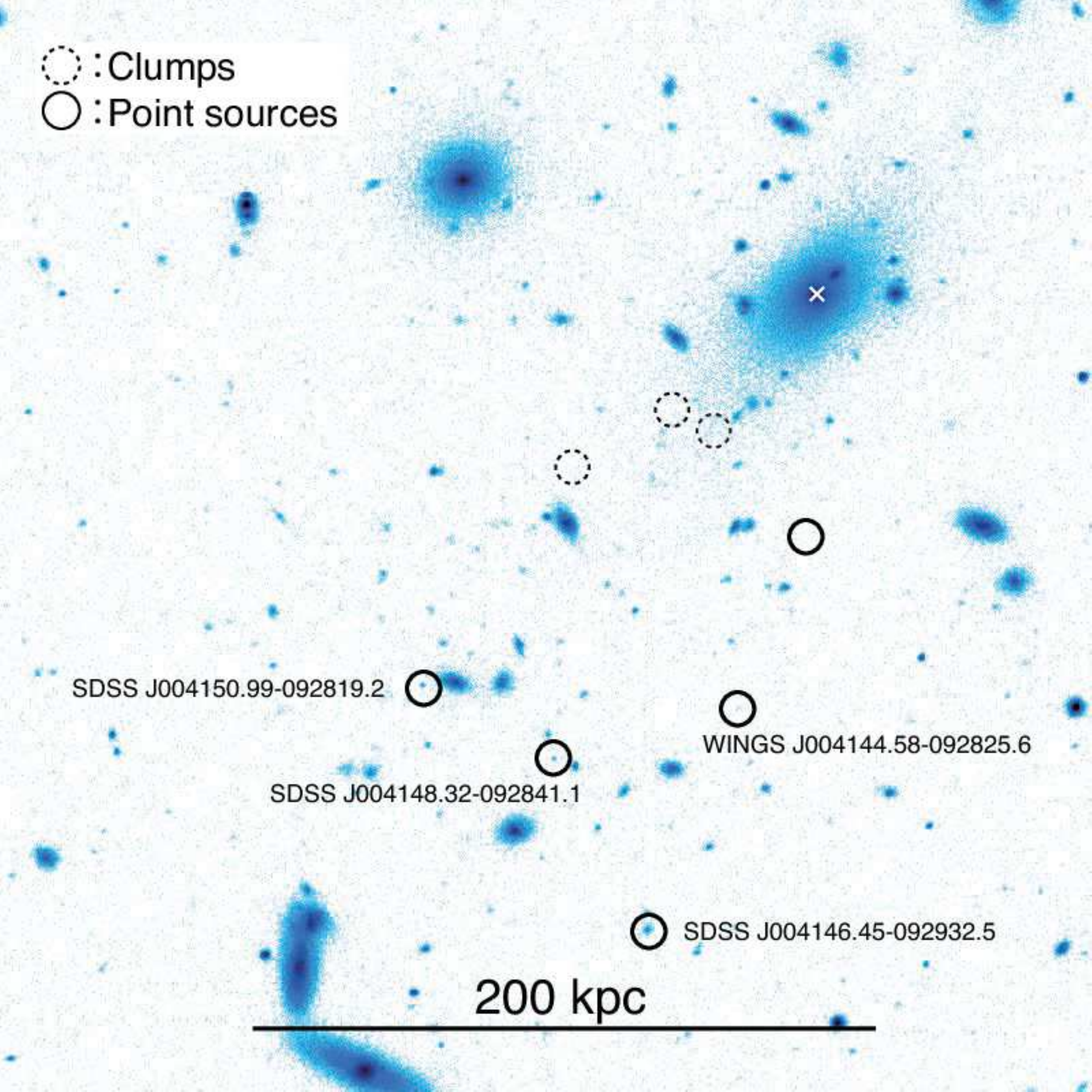}
  \end{center}
 \end{minipage}
 \caption{{\it Left}: Closeup {\it Chandra} image of the S subcluster ($\sigma=4.0~\mr{arcsec}$ Gaussian smoothed). {\it Middle}: {\it Chandra} relative deviation image of the same sky region ($\sigma=6.8~\mr{arcsec}$ Gaussian smoothed). {\it Right}: SDSS r-band optical image of the same sky region. The cross represents the position of the brightest central galaxy of this subcluster. The clumps in the subcluster tail and the nearby point sources are denoted with dashed and solid circles, respectively. }
 \label{img_subcluster}
\end{figure*}

Fig. \ref{img_subcluster} shows a closeup view of the S subcluster. We see a contact discontinuity (brightness edge) at the north of its core. The X-ray brightness peak of the subcluster is spatially coincident with the position of its central dominant galaxy. The core appears extended to the south. From the core of the subcluster, a bright tail extends in the southeastern direction. The north edge of the tail extends from the core to the east for 200~kpc, and appears remarkably straight and smooth. In contrast, the southwestern edge appears blurred, indicating that the ICM transport properties are different along these two edges. About 150~kpc southeast of the core, the tail appears bent (see also the thermodynamic maps in Fig. \ref{map}). The tail shows an abrupt brightness drop at $\sim$200~kpc from the core.

The thermodynamic maps (see Fig. \ref{map} and Fig. \ref{map_tdiv}) show that the subcluster has a low-temperature, low-entropy core from which the broad low-entropy gas tail extends in the southeastern direction. The pressure of the S subcluster gas is higher than that of other regions at the same distances from the main cluster core. The radial change of the pressure along a line from the main cluster core to the S subcluster is gradual while the temperature, the density and the entropy profiles change abruptly at the brightness edge of the S subcluster.

Furthermore, a region of high-temperature and high-entropy gas, a ``hotspot'',  is seen between the subcluster and the main cluster, to the northeast of the subcluster core \citep[see also][]{tanaka10}.

\subsubsection{Clumpy gas and the SE brightness excess}\label{clumpygas}
 In the {\it Chandra} closeup image (Fig. \ref{img_subcluster}), we see three gas clump candidates - X-ray bright sources in the tail with no optical counterparts within a 3~arcsec radius. They have been detected with the \verb+wavdetect+ tool (along with the point sources, see Section \ref{thermodynamic_mapping}) at a higher than 3$\sigma$ significance.

\begin{figure}
\begin{center}
\includegraphics[width=3.4in]{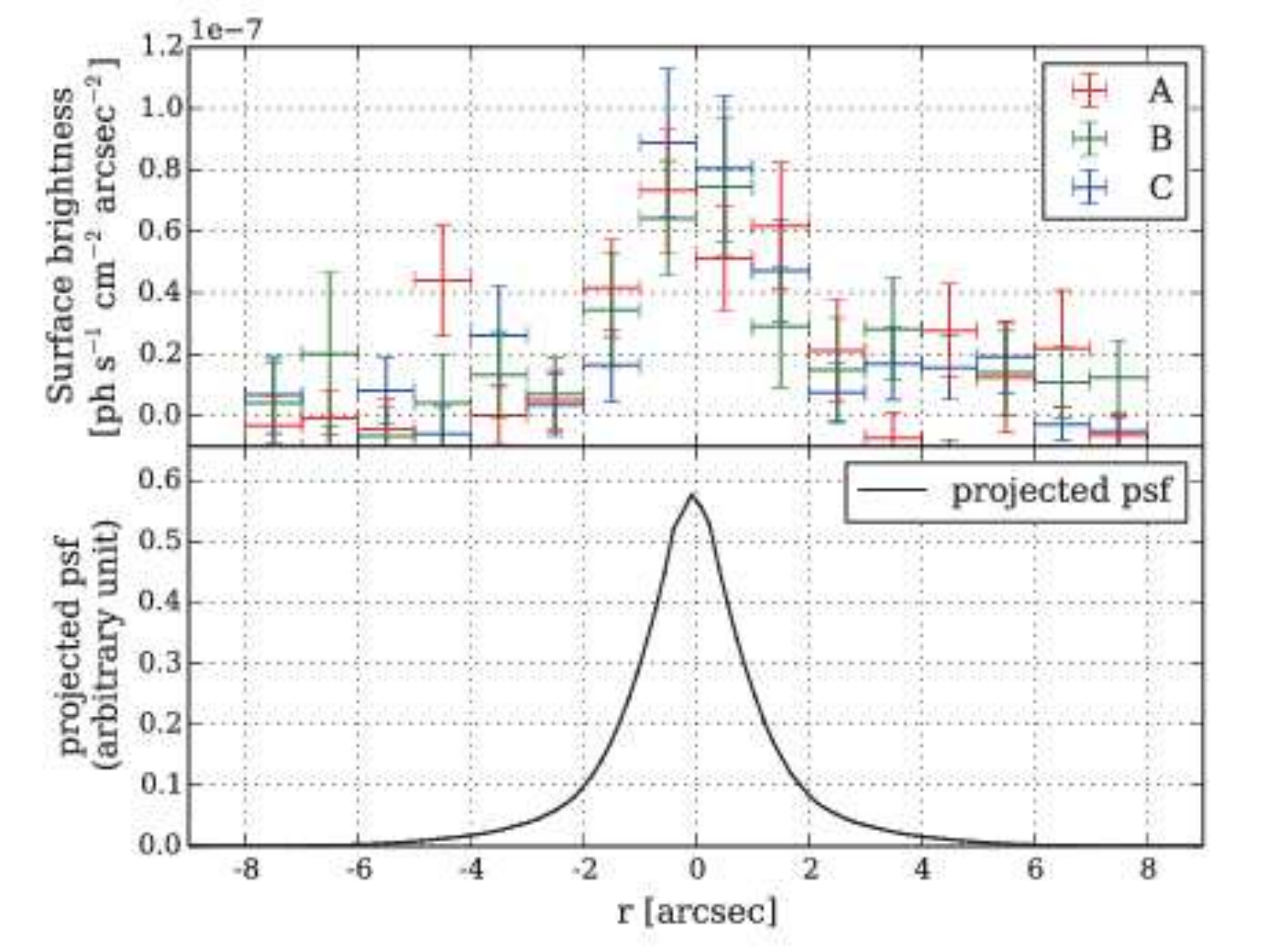}
\end{center}
\caption{{\it Top}: Surface brightness profiles of the clumps shown in Fig. \ref{img_subcluster}. {\it Bottom}: Surface brightness profile (projected psf including multiple observations) of an arbitrary nearby X-ray point source calculated using the {\small CIAO} \texttt{psf} module.}
\label{clump}
\end{figure}

Fig. \ref{clump} top shows the surface brightness profiles (projected for 4~arcsec) across the clump candidates. The projected profile of each clump candiate is well fitted with a Cauchy plus linear function. The FWHM of all three clump candidates calculated using the widths of the best-fitting functions are in the range of 2.1--2.8 arcsec for the Cauchy profile, which is about $1\sigma$ larger than the PSF model or the FWHM of the nearby point sources shown in Fig.~\ref{img_subcluster}, except source `d'. The luminosities of the clumps are $\sim1$--$2\times10^{40}$~erg~s$^{-1}$.

As seen in Fig. \ref{img_xmm},  a surface brightness excess extends from the tail of the S subcluster all the way out to $\sim r_{500}$ (SE brightness excess). This brightness excess has been previously reported based on both {\it ROSAT} and {\it XMM-Newton} observations \citep{durret98,durret03,durret05a}.

\subsubsection{Deprojected profile out to the outskirts}
\begin{figure}
\begin{center}
\includegraphics[width=3.4in]{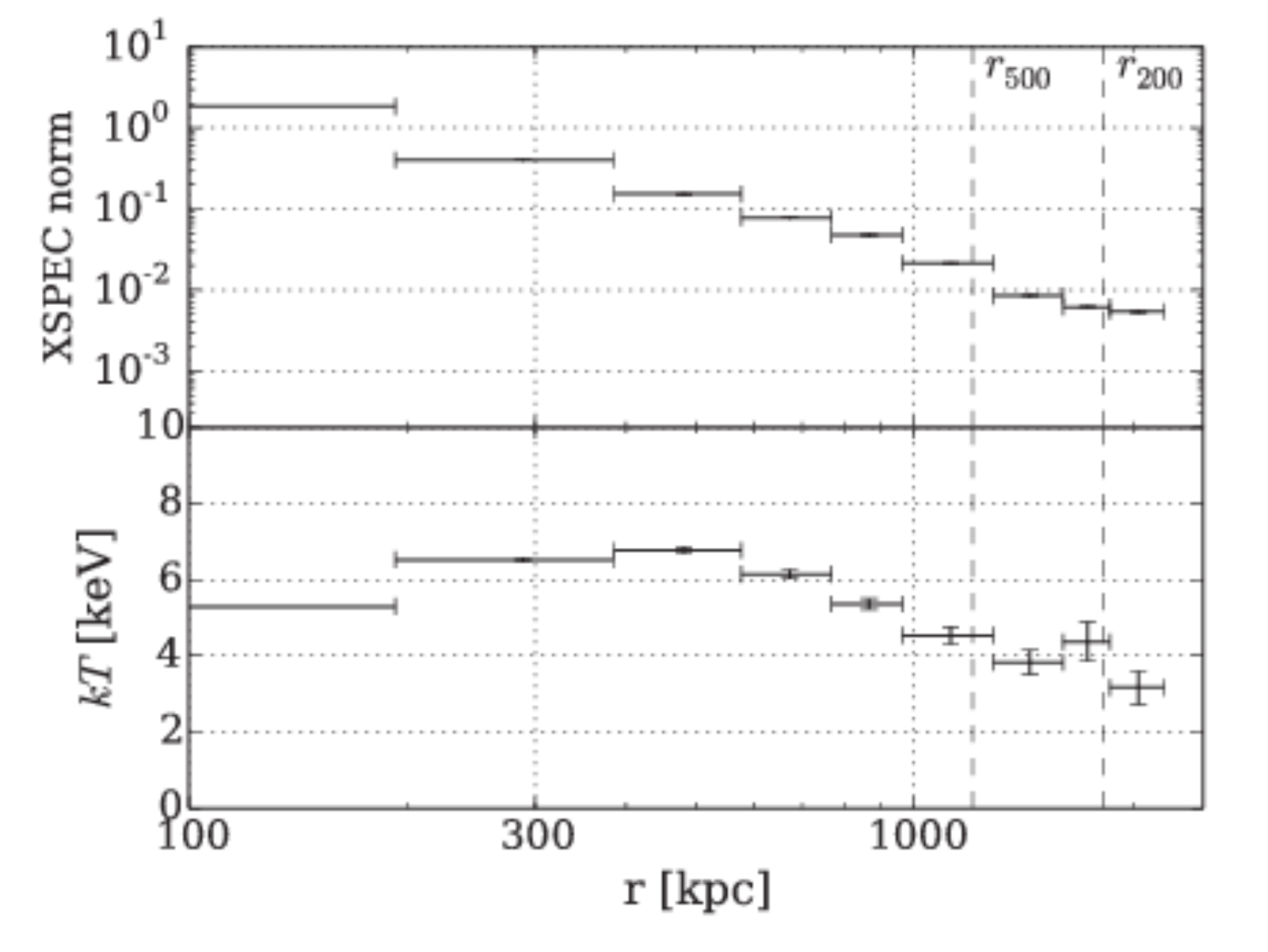}
\end{center}
\caption{{\it Suzaku} projected radial profiles. The panels are {\small XSPEC} normalization per unit area and temperature from top to bottom.}
\label{proj}
\end{figure}
\begin{figure}
\begin{center}
\includegraphics[width=3.4in]{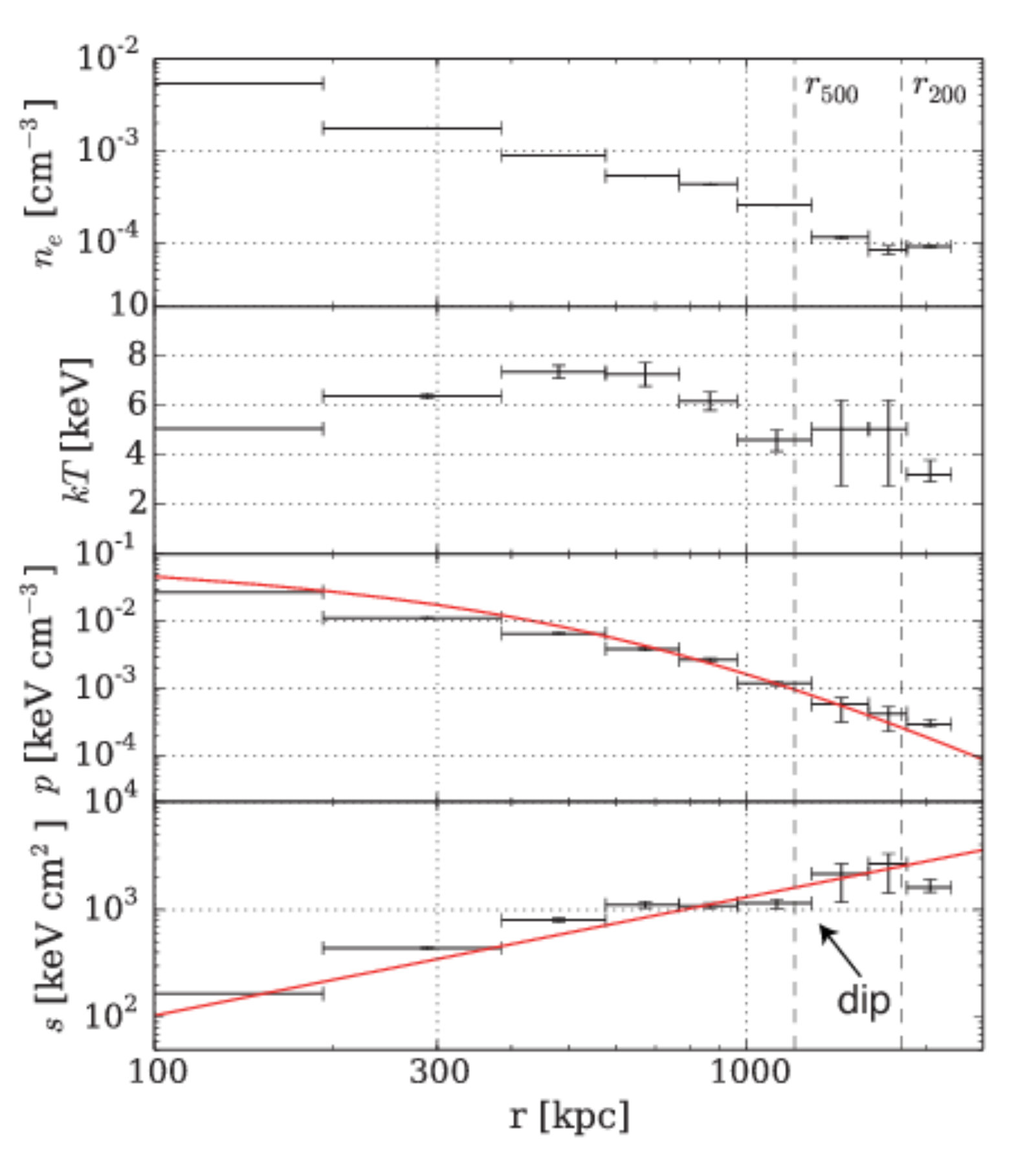}
\end{center}
\caption{{\it Suzaku} deprojected radial thermodynamic profiles. The panels are density, temperature, pressure and entropy from top to bottom. The curve on the pressure profile is the best-fitting theoretical pressure profile \citep{nagai07}, with the parameters estimated in \citet{planck13}. The line on the entropy profile is the theoretically predicted entropy profile \citep{pratt10}.}
\label{deproj}
\end{figure}

In Fig. \ref{proj} and \ref{deproj}, we present the projected and deprojected thermodynamic profiles measured along the direction of the infall of the S subcluster using the {\it Suzaku} data. The low-entropy S subcluster has been excised from the {\it Suzaku} data analysis, otherwise, it would have dominated the ICM signal at those radii, and would have caused a severe departure from spherical symmetry making the deprojection analysis practically meaningless. The curves in the deprojected pressure and entropy profiles are the theoretically predicted profiles calculated following \citet{nagai07} and \citet{pratt10}, respectively. We fitted the pressure profile with the model leaving $r_{500}$ as a free parameter with other parameters fixed to the values presented in \citet{planck13}. $M_{500}$ is expressed self-consistently as a cubic function of $r_{500}$, with the normalization of $M_{500}/{r_{500}}^3 = 7.2\times 10^{14}\mr{M_\odot}/(1.33~\mr{Mpc})^3$ \citep{mantz10a, mantz10b}. The best-fitting value of $r_{500}=1.2\pm0.04$~Mpc and the corresponding $M_{500}$ are also used in the entropy model.

As seen in both the projected and deprojected profiles, the temperature remains above $\sim$4~keV out to $r_{200}$. The deprojected density decreases monotonically to $r_{200}$, but flattens out in the outermost data point that lies beyond $r_{200}$. We note that we have used an extrapolation of the best-fitting beta model for the density (obtained ignoring the innermost two data points representing the cool core) in order to correct the density profile for the effect of projected emission from gas beyond the outer edge of the measurements.

The deprojected entropy profile shows an expected low-level excess above the expected power-law behaviour near the cluster core, which is attributable to additional heating due to the AGN feedback and ongoing mergers; at larger radii, the measured entropy agrees with the model out to $r_{200}$, with the exception of a dip found around 1.3~Mpc ($\sim r_{500}$) and which is probably associated with the low-surface brightness extension of the tail of the S subcluster.

Previous studies \citep[e.g.][]{simionescu11,walker13,morandi13,urban14} have revealed a flattening of the measured entropy profile in other clusters with respect to the expected model, starting at around $\sim$0.75$r_{200}$. In the case of Abell 85, by contrast, the deprojected entropy profile remains in good agreement with the model until $r_{200}$ and turns over only beyond that radius. However, the large error bars on the last two data points inside $r_{200}$ do not rule out a flattening of the entropy in Abell 85 in agreement with that observed in other systems. Moreover, the outermost data point in our profile shows an unusually high density (in most other systems, the density profile is monotonically decreasing even beyond $r_{200}$). If therefore the density in this outermost data point is biased high, either by the presence of an unresolved group or due to an unusually high clumping factor that is not azimuthally representative, the measured entropy immediately inside $r_{200}$ could be biased high as a consequence.

\subsection{Main cluster features}
The main cluster has an asymmetric surface brightness morphology in which the core gas extends farther towards the northern direction whereas the outer gas extends south (Fig. \ref{img_chandra} left).
\subsubsection{Large scale spiral}
In the {\it Chandra} relative deviation image (Fig. \ref{residual_image} left), we see an apparent brightness excess spiral, starting north of the core and extending counter-clockwise outward from the core. In the {\it XMM-Newton} relative deviation image (Fig. \ref{residual_image} right), we can see that this spiral extends out to $\sim$600~kpc. This indicates ongoing gas sloshing, and the appearance of the spiral feature suggests that  the interaction plane is close to the plane of the sky \citep{roediger11}. Sloshing motions extending to large radii have previously been seen in the Perseus Cluster \citep{simionescu12}, Abell 2142 \citep{rossetti13}, Abell 2029 \citep{paterno-mahler13} and RXJ 2014.8-2430 \citep{walker14}.

In the trend-divided density map (Fig. \ref{map_tdiv}), an obvious counter-clockwise spiral feature of higher density gas can be seen around the main cluster core, corresponding to a sloshing spiral in the relative deviation images. Indications for this spiral structure are also seen in the temperature and entropy maps but cannot be seen in the pressure map. Instead of the spiral feature, the trend-divided pressure map displays an asymmetric morphology with a pressure excess in the southern and southeastern direction from the main cluster core.

\subsubsection{Main cluster core}
\begin{figure}
 \begin{center}
  \includegraphics[width=1.6in]{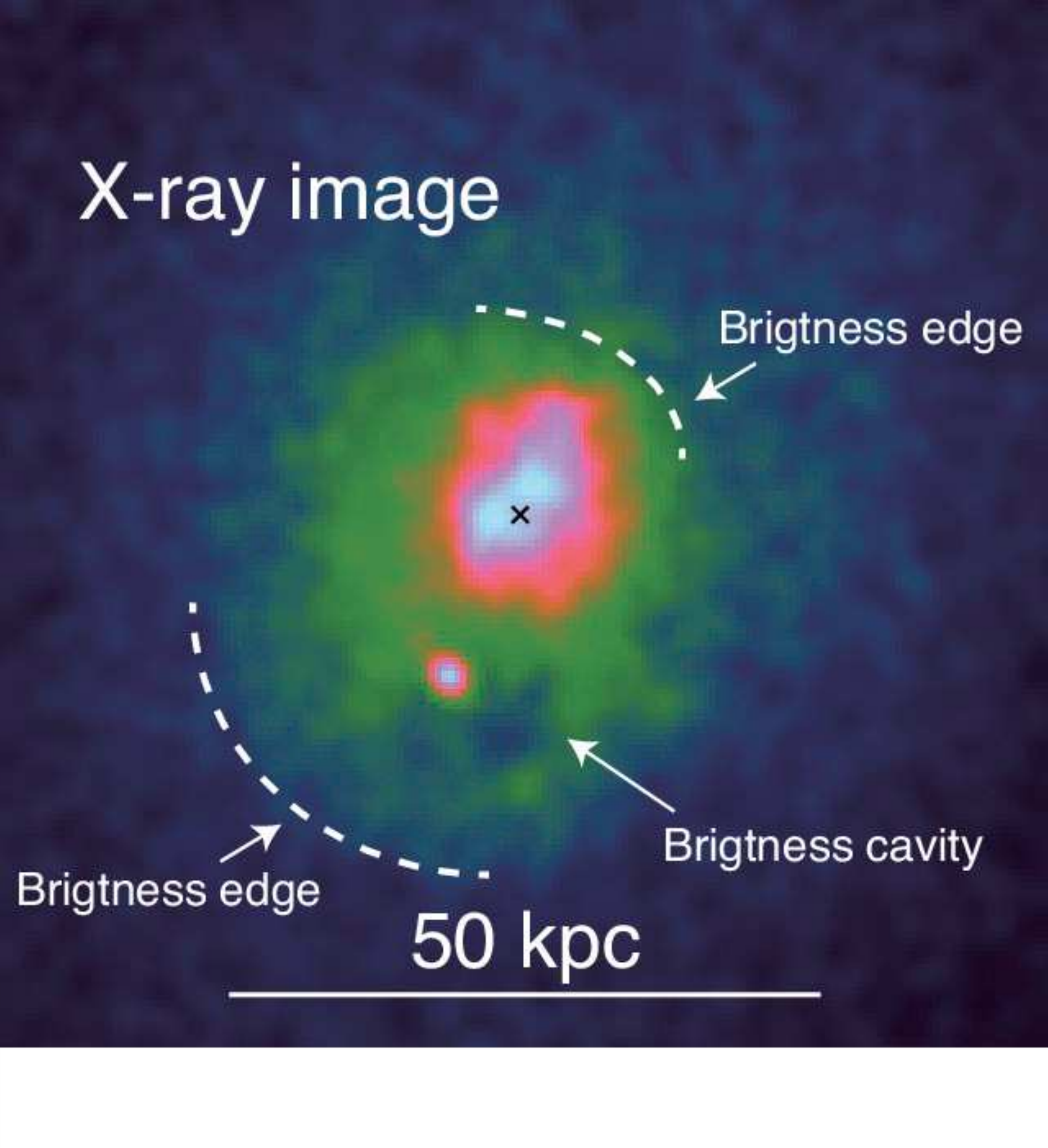}
  \includegraphics[width=1.6in]{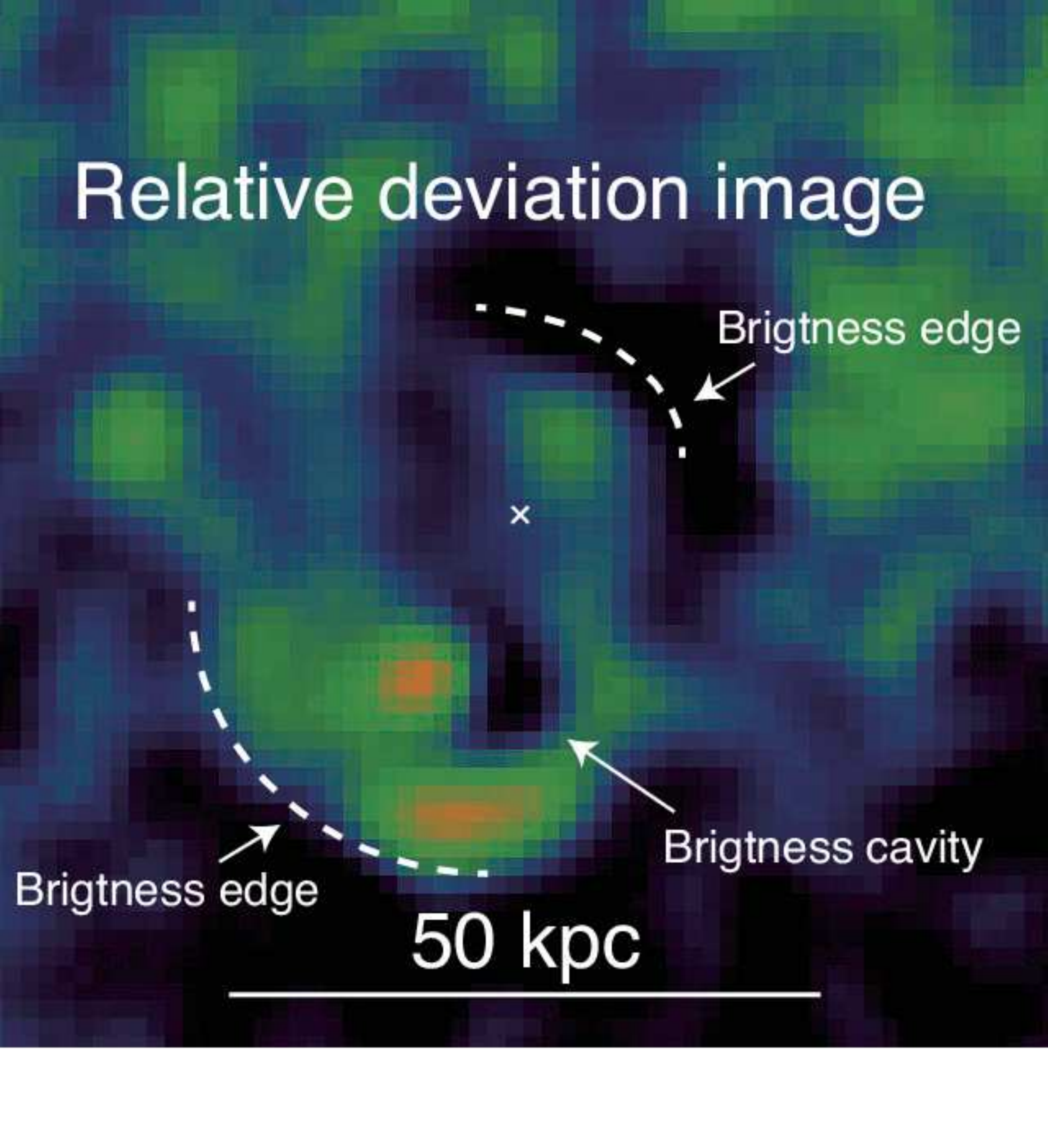}
 \end{center}
\caption{{\it Left}: Closeup image of the main cluster core($\sigma=2.0~\mr{arcsec}$ Gaussian smoothed). {\it Right}: Relative deviation image of the same region (see also Fig. \ref{residual_image}). The cross corresponds to the central cD galaxy.}
 \label{img_core}
\end{figure}

Closeup images (an X-ray image and a relative deviation image) of the main cluster core are shown in Fig. \ref{img_core}. We see a surface-brightness cavity at 20~kpc south of the core. In addition to the cavity, we see two arc-shaped brightness edges in the image. One is located northwest of the brightness peak, while the other is at $r=35$~kpc to the south.

\subsection{SW subcluster and the dark band}
The temperature and entropy of the SW subcluster are relatively low, but it does not have a cool core (it has no surface brightness peak or an obvious central dominant galaxy). It has a diffuse, low-temperature and low-entropy tail extending to the west, and its pressure is consistent with that of the surrounding gas  (Fig. \ref{map}).

About 250~kpc southwest of the main cluster core, at the interface between the sloshing spiral and the SW subcluster, we see a straight, dark narrow structure (Dark band) with a width of $\sim$50~kpc and a length of $\sim$300~kpc which divides the SW subcluster from the sloshing gas of the main cluster (Fig. \ref{residual_image}). The part of the spiral adjacent to the Dark band is brighter than the other parts of the spiral. The regions corresponding to the Dark band have relatively high temperatures (Fig. \ref{map}).

\section{Discussion}\label{discussion}
\subsection{The infall of the S subcluster}
\subsubsection{Dynamics of the S subcluster}
In the temperature map (Fig. \ref{map} upper right), we see a high-temperature region (hotspot) in the northeast of the S subcluster. This hotspot has been first observed by {\it XMM-Newton} \citep{durret05b}. \citet{tanaka10}, who conducted a subsequent {\it Suzaku} observation, proposed a scenario that the hot gas results from the merger of the S subcluster, colliding with the main cluster from the southwest.

In our thermodynamic maps (Fig. \ref{map}), the low entropy tail of the subcluster extends to the southeast. It is thus more likely that the subcluster is moving northwestward rather than northeastward as proposed in \citet{tanaka10}.  The tail comprises two gas components of different entropy: higher entropy ($\sim$800 keV cm$^{2}\times(l/1~{\rm Mpc})^{1/3}$; green bins associated with the S subcluster in the bottom right panel of Fig.~\ref{map}) gas extending broadly, and lower entropy ($\sim$400 keV cm$^{2}\times(l/1~{\rm Mpc})^{1/3}$; blue bins associated with the S subcluster in the bottom right panel of Fig.~\ref{map}) gas terminating at $\sim$200~kpc from the subcluster's core, similar to what we see in the X-ray image (Fig. \ref{img_subcluster}). While the relatively higher entropy gas may have been stripped from the outer X-ray halo of the S subcluster, the dense low-entropy gas is likely being stripped from its cool core remnant. The emissivity of the stripped higher entropy gas is significantly lower than that of the low entropy tail, resulting in an abrupt drop in surface brightness at the end of the tail.

\subsubsection{Destruction of the cool core}\label{dest_cc}
Because the innermost low-entropy core of the subcluster along with its brightest central galaxy precedes the stripped tail, the cool core of the subcluster has likely been almost completely destroyed before reaching its current radius, $r\approx500$~kpc of Abell~85. The possible gas clumps that we identified within the low entropy tail have no optical counterparts. Their luminosities of $L_{\rm X}=1$--$2\times10^{40}$~erg~s$^{-1}$ are consistent with the luminosities of the coronae of late type galaxies \citep{sun2007}, indicating that they might have been stripped off the member galaxies of the infalling group.

\begin{figure}
 \begin{center}
  \includegraphics[width=3.3in]{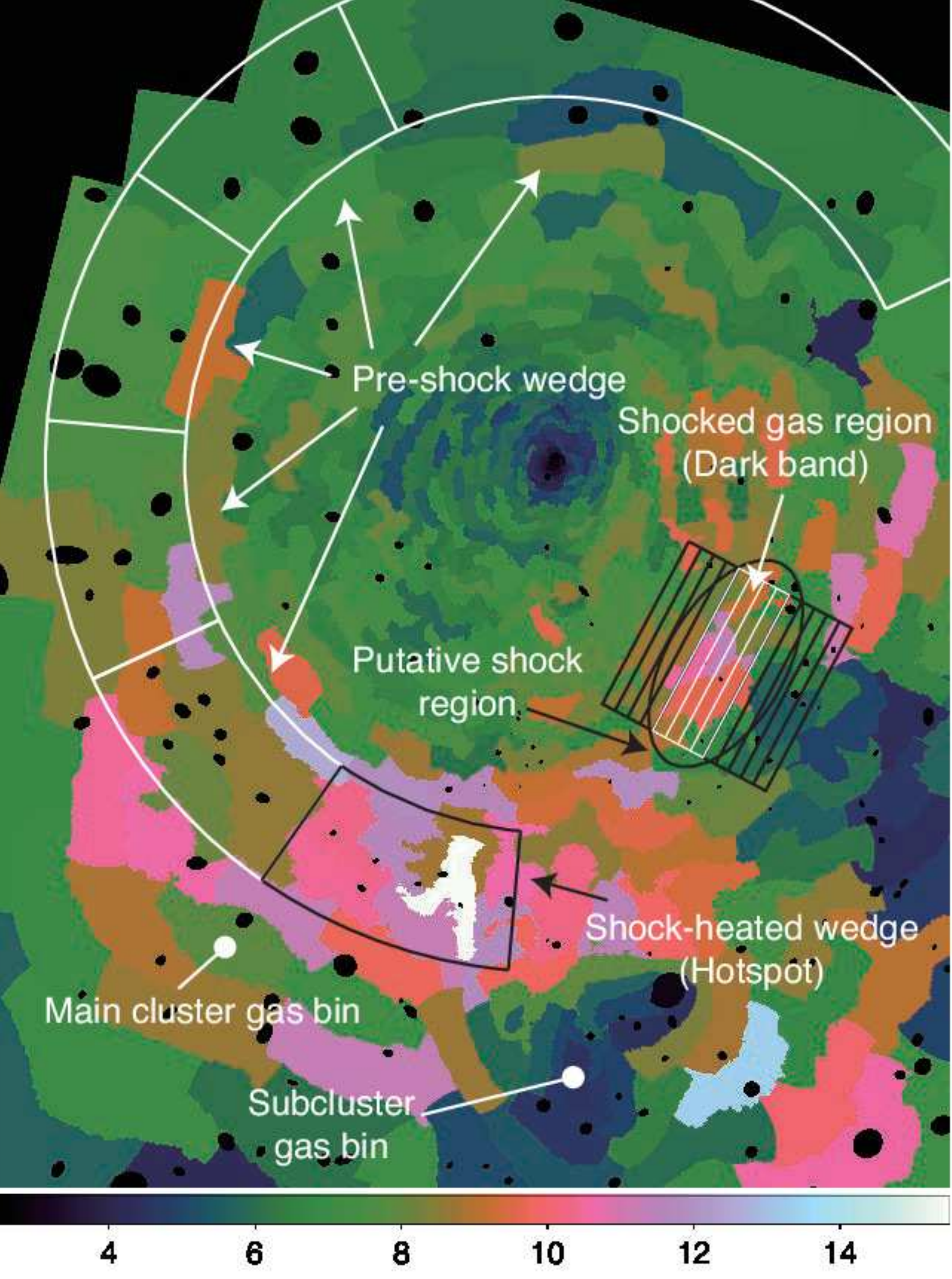}
 \end{center}
 \caption{Closeup image of the temperature map. Two bins (Main cluster gas bin and Subcluster gas bin) are used to estimate the projection effect of the subcluster tail gas in Section \ref{dest_cc}. The bottommost black wedge (Shock-heated wedge) is used to extract the properties of the shock-heated gas and the white wedges (Pre-shock wedge) are used to evaluate the pre-shock gas properties in Section \ref{hotspot}. The overlaid rectangular regions are used to extract the thermodynamic profiles in Section \ref{sw}. The black ellipse denotes the putative shock region, and the white rectangles correspond to the Dark band.
}
 \label{hotspot_regions}\label{img_shock}
\end{figure}

Excluding the immediate surrounding of the brightest central galaxy of the merging subcluster, its projected temperature is $\sim$4.5~keV. However, since this value is the emission weighted average of the ambient ICM and the stripped tail, the temperature is probably overestimated. To estimate the effect, we chose two bins representing the cluster ambient gas and the subcluster tail gas as shown in Fig. \ref{hotspot_regions}, and fitted the spectra for the subcluster tail bin using the main cluster gas spectra as background. The spectrum of the main cluster gas bin was fitted with a single temperature plasma model. The temperature obtained for the main cluster gas bin was used as one of the temperature components of the two temperature plasma model with which we fitted the spectrum of the subcluster tail gas bin.

The best-fitting temperature of the tail changes from $kT = 4.5^{+0.4}_{-0.3}$~keV with the single temperature model to $kT = 2.7^{+0.5}_{-0.4}$~keV with the two temperature model. Under the assumption that the stripped tail gas did not mix with and did not get conductively heated by the ambient medium, we can use this temperature to estimate the mass of the infalling subcluster. For the 2.7~keV temperature, the mass-temperature scaling relation by \citet{arnaud05} predicts $M_{500} \approx $1.3$\times 10^{14}\mr{M_\odot}$. \citet{mantz10b} estimated the main cluster mass as $M_{500} \approx 7.2 \times 10^{14}\mr{M_\odot}$, which implies a merger mass ratio of $\sim5.5$.

Since this 2.7~keV temperature is extracted for the vicinity of the cool core region of the subcluster and this temperature is not necessarily representative of the subcluster as a whole, the mass of the subcluster may be underestimated. However, regardless of the exact mass ratio, this demonstrates that cool cores can be effectively stripped during mergers. Recently, observations of the Ophiuchus Cluster have shown that sloshing induced by major late time mergers may even destroy the cool core of the main cluster \citep{million10}. These observations pose a challenge to simulations \citep[e.g.][]{burns08}, which find that cool cores usually survive late major mergers.

\subsubsection{Stripped tail and gas clumping}\label{tail}
{\it ROSAT} and {\it XMM-Newton} images of Abell~85 \citep{durret98,durret03,durret05a} have shown a large-scale brightness excess structure extending to the southeast of the S subcluster. \citet{durret03} estimated the temperature of this structure as $\sim$2~keV and proposed that it may be either due to the diffuse emission of a large scale structure filament connecting to the cluster, due to a chain of small groups of galaxies, or due to the stripped gas from the infalling S subcluster. \citet{boue08} conducted an optical study and concluded that the structure is consistent with groups falling into the main cluster.

However, the temperature profile measured by {\it Suzaku} shows $kT>4$~keV out to $r_{200}$. This temperature is a factor of two higher than the previous measurement of $\sim$2~keV at the same radius. This discrepancy may be the result of the difference in the background treatment. While \citet{durret03} used blank-sky templates, we determined our local background using the outermost regions of the {\it Suzaku} observation. We find that the Galactic foreground component toward Abell~85 is a factor of three higher than the average value across the sky \citep{kuntz00}. Using background templates will therefore underestimate the background and lead to a lower best-fitting temperature in the low surface brightness cluster outskirts.

If most of the excess emission were due to groups, we would expect $kT \approx 1$~keV for a typical group mass. The {\it Suzaku} spectra are inconsistent with such low temperature emission and they show that the temperature of the southern surface brightness excess is consistent with the projected temperature of the bright  stripped tail of the S subcluster observed by {\it Chandra}. It is thus likely that the SE brightness excess is also associated with the stripped gas of the infalling S subcluster. Long tails of stripped gas that survive within the ambient ICM for over 600 Myr have recently been detected in the outskirts of the cluster Abell~2142 \citep{eckert2014} and in the Coma Cluster \citep{sanders2013}.

This interpretation is also supported by the dip in the deprojected entropy profile. The location of this dip at $r_{500}$ corresponds to the apparent termination of the SE brightness excess. The dip, therefore, may be caused by the projection of the low-entropy, bright tail gas against a background of high-entropy, faint main cluster gas. Note that this interpretation does not contradict the galaxy overdensity found in the southeast \citep{boue08} because cluster mergers usually happen along large-scale filaments (in the cosmological sense), and we expect more galaxies along filaments than other directions.

The fact that the stripped tail of the infalling southern subcluster in Abell~85 is seen across a radial range of over 700~kpc, as well as the recent observations of the long trails of stripped gas in other systems \citep{sanders2013,eckert2014},  indicate that the stripping of infalling subclusters may seed gas inhomogeneities in the outskirts of clusters. The presence of gas clumping in cluster outskirts is seen in simulations \citep[e.g.][]{roncarelli06,roncarelli2013,nagai11,vazza2013}, and X-ray observations of clusters out to their virial radii have indirectly shown that clumping becomes significant at $r > 0.5r_{200}$ \citep{simionescu11, walker13, morandi13,urban14}. In the case of Abell 85, the deprojected entropy profile remains in good agreement with the expected power-law model until $r_{200}$, which would imply, at face value, that in this system clumping is not significant at radii  $r_{500}<r<r_{200}$. However, this conclusion cannot be drawn robustly because the large error bars on the last two data points inside $r_{200}$ do not rule out a flattening of the entropy in Abell 85 in agreement with that observed in other systems. Moreover, this profile is extracted along the path of the S subcluster, which is clearly an unrelaxed direction. This may lead to the break of the assumption of the spherical symmetry in deprojection.

\subsection{Gas motion and interaction}

\subsubsection{Gas sloshing in the main cluster}\label{sloshing}
\begin{figure*}
 \begin{minipage}{0.495\hsize}
  \begin{center}
   \includegraphics[width=3.5in]{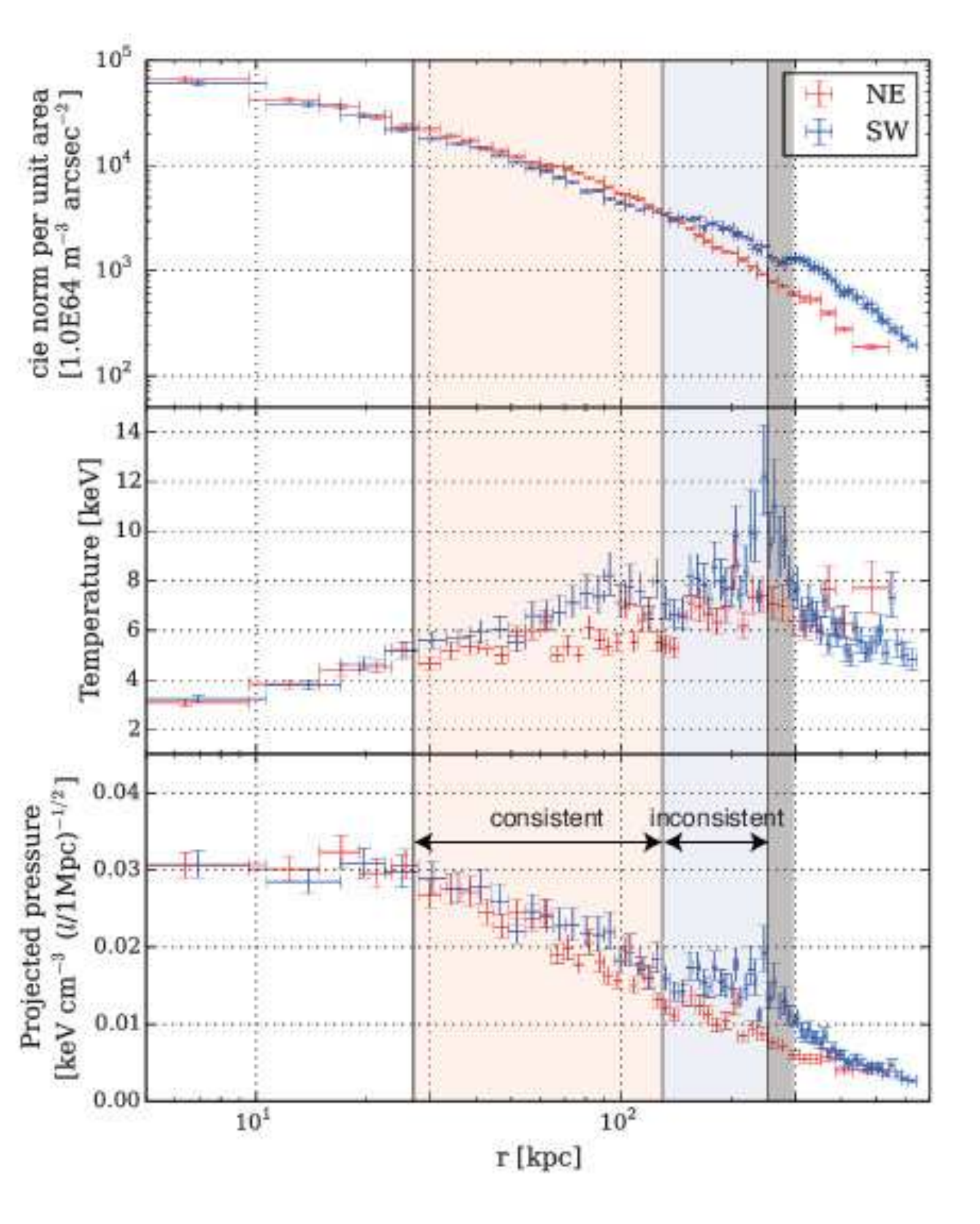}
   \includegraphics[width=3.5in]{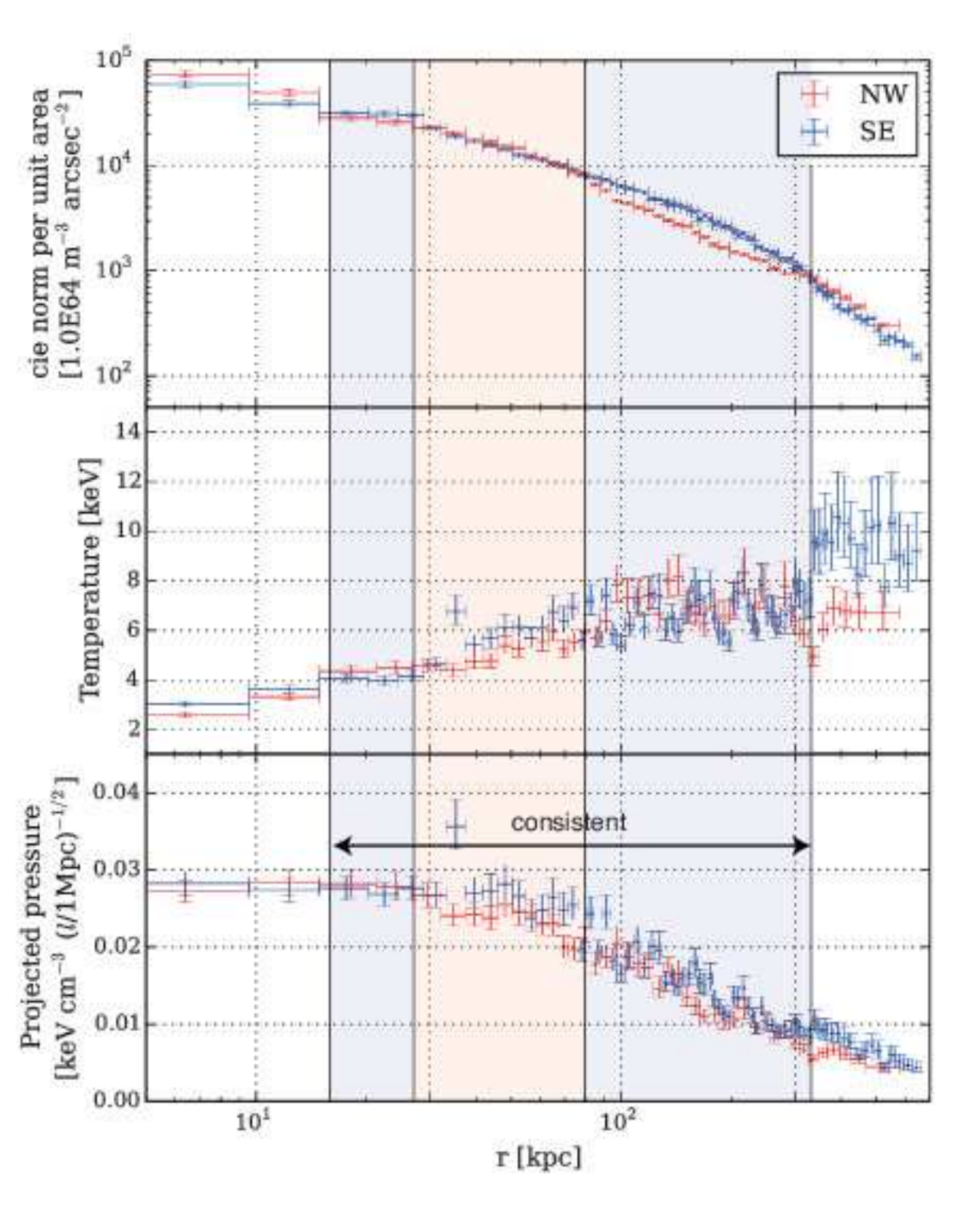}
  \end{center}
 \end{minipage}
 \begin{minipage}{0.495\hsize}
  \begin{center}
   \includegraphics[width=3.5in]{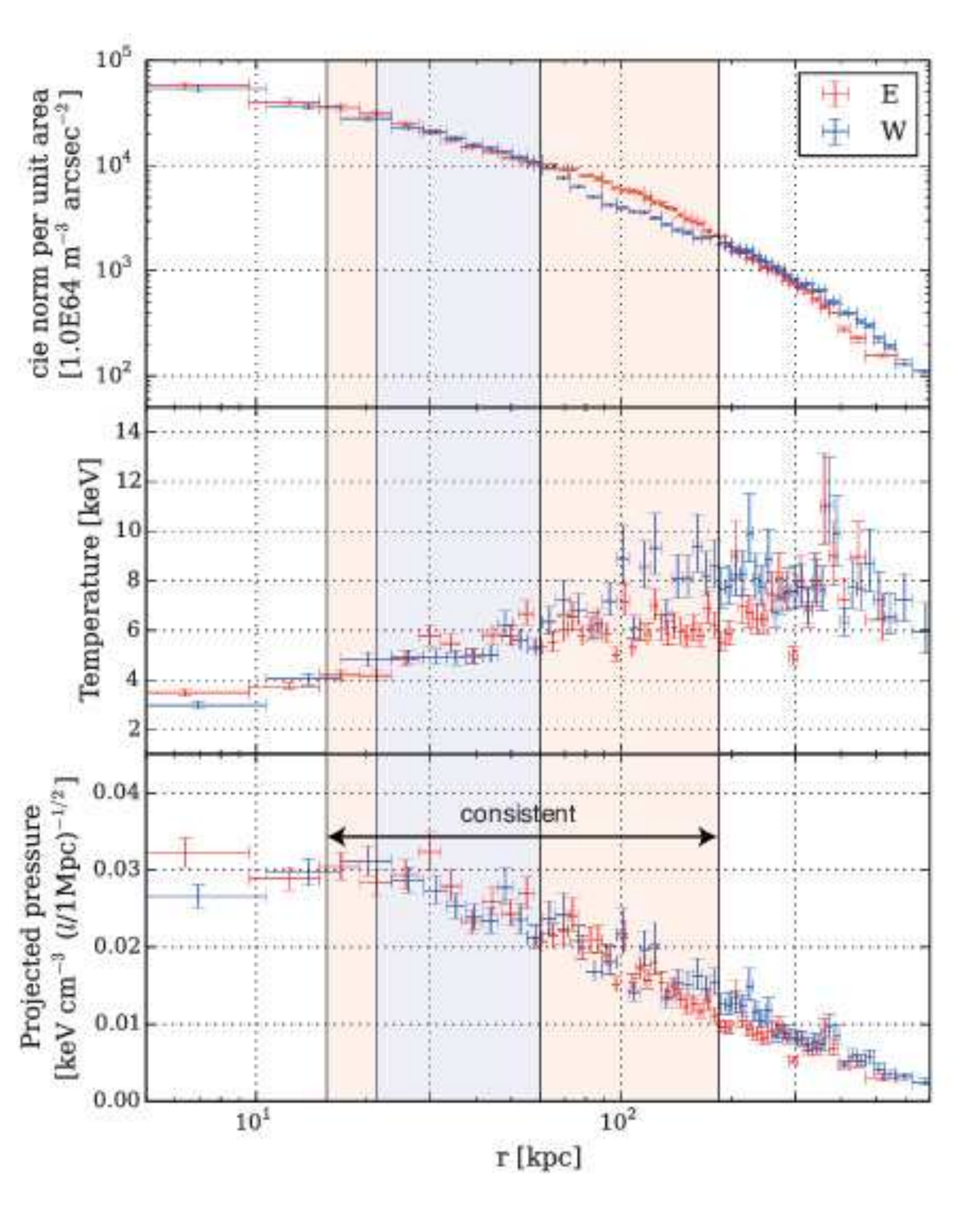}
   \includegraphics[width=3.5in]{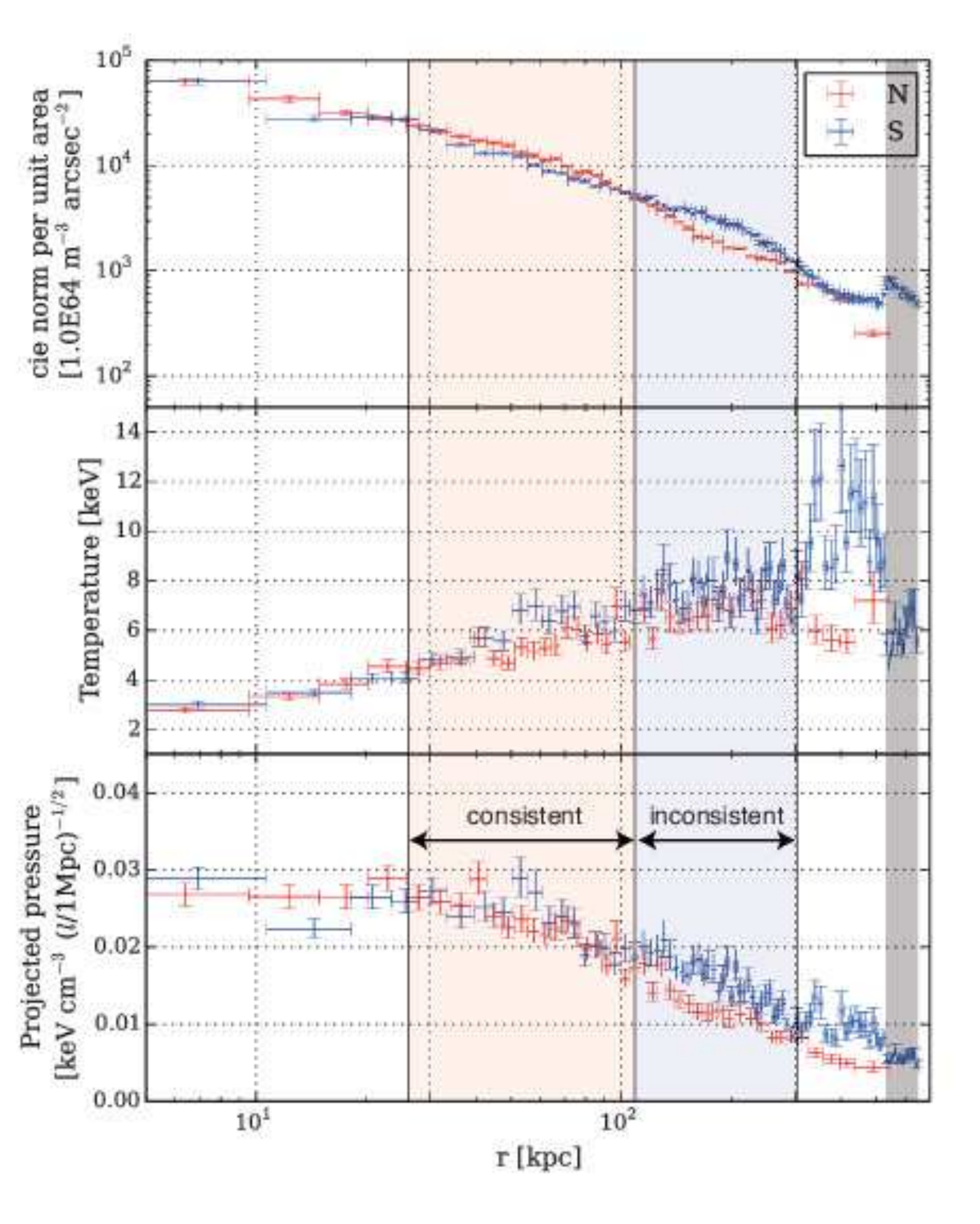}
  \end{center}
 \end{minipage}
 \cprotect\caption{Radial projeced thermodynamic profiles for eight directions illustrated in Fig. \ref{residual_image} left. Two profiles for an opposite direction pair are drawn in each panel. Individual panels are surface brightness (\verb+cie+ normalization per unit area), temperature and pressure, respectively from top to bottom. The black vertical lines represent the approximate locations of the spiral edges. The red and blue rectangles correspond to the spiral arm, whose colors correspond to the color of the brighter direction of the two directions in the relative deviation images (Fig. \ref{residual_image}). The grey rectangles in the upper left panel and bottom right panel represent the Dark band illustrated in Fig. \ref{residual_image} right and the S subcluster, respectively.}
 \label{sloshing_profile}
\end{figure*}

The images and thermodynamic maps of Abell~85 show spiral like features indicative of gas sloshing (see Fig. \ref{residual_image} and Fig. \ref{map_tdiv}). For detailed simulations and discussion on gas sloshing, see e.g. \citet{tittley05}, \citet{ascasibar06}, \citet{roediger11}, \citet{zuhone11}.

In Fig. \ref{sloshing_profile}, we display the projected radial profiles for the surface brightness and several thermodynamic quantities for the eight directions shown in Fig. \ref{residual_image}. For comparison, two profiles of an opposite direction pair (e.g. N and S) are drawn in each panel. The overlaid red/blue rectangles represent the radial intervals where the profile drawn in the corresponding color is brighter in the relative deviation images.

For most of the regions, the surface brightness profiles and the temperature profiles show an anticorrelation typical of sloshing phenomena, namely, alternating bright cooler regions and dim hotter regions \citep[see e.g.][]{tittley05}. The resulting pressure profiles, each of which is a simple product of density and temperature, are in these regions consistent between the two opposite directions (see annotations in Fig. \ref{sloshing_profile}).
However, beyond $\sim$120~kpc of the NE-SW and the N-S directions (the blue rectangles in the NE-SW and N-S profiles), the surface brightness profiles corresponding to the outermost part of the spiral arm are brighter than those of the opposite directions, while the temperature profiles are either similar or the brighter part is hotter, resulting in a clear discrepancy in the pressure profiles between two opposite directions (see also Fig. \ref{map_tdiv}).

The high-pressure gas is most likely shock-heated and compressed by the infalling two subclusters. Both subclusters have associated putative shock-heated regions (see Section \ref{shock}) and we can expect shock-heated gas anywhere around the region of interaction. The fact that these two subclusters are both located in the southwestern half of the main cluster is consistent with the higher pressure in the southwestern part of the spiral.

\subsubsection{Gas sloshing and the morphology of the S subcluster}\label{interaction}
\begin{figure}
\begin{center}
\includegraphics[width=3.5in]{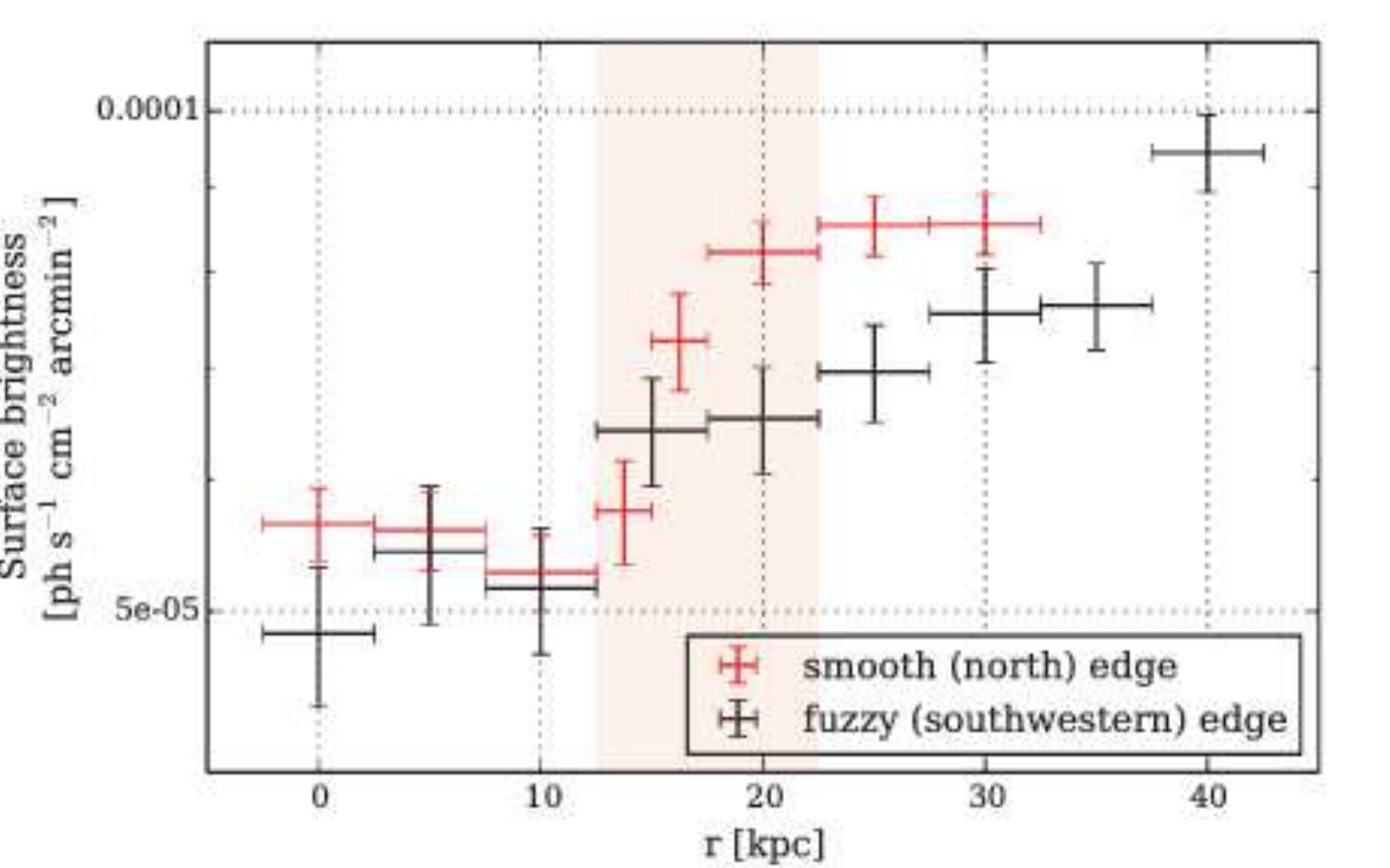}
\end{center}
\caption{Surface brightness profiles across two edges shown in Fig. \ref{img_subcluster}. The red and black points correspond to the north edge and the southwestern edge respectively. The profiles are drawn with the directions shown by the arrows in Fig. \ref{img_subcluster}. The red rectangle denotes the north edge.}
\label{edge}
\end{figure}

The bright stripped tail of the S subcluster shows a bent structure and has a 200~kpc long, relatively smooth and sharp northern edge, as described in Section \ref{ssub}. Fig. \ref{edge} shows the surface brightness profiles across the northern and southwestern edge. While the brightness of the southwestern edge changes gradually, the profile across the north edge is steep, with a width of $\sim$10~kpc.

The Coulomb mean free path $\lambda$ of electrons in the ambient medium satisfies the equation \citep{markevitch07}
\begin{eqnarray}
 \lambda = 15~\mr{kpc}\left(\dfrac{kT}{7~\mr{keV}}\right)^2\left(\dfrac{n_{\rm e}}{10^{-3}~\mr{cm}^{-3}}\right)^{-1},
\end{eqnarray}
where $n_{\rm e}$ is the electron density. The typical temperature and density of the ambient ICM around the northern edge are $\sim$10~keV and $\sim$10$^{-3}$~cm$^{-3}$, respectively, so the Coulomb mean free path is estimated as $\lambda \approx 30$~kpc. This value is larger than the width of the northern edge of $\sim$10~kpc, indicating suppression of transport processes.

Ordered flows of the ambient ICM from west to east, resulting from sloshing, can consistently explain both the smooth brightness edge and the bent morphology. It has been shown in simulations that sloshing causes ordered gas motion, and that magnetic fields are stretched and regulated by the velocity field \citep{ascasibar06,roediger11,zuhone11,zuhone14}. As the sloshing spiral in Abell~85 extends out to $\sim$600~kpc, there may be a non-negligible flow around the S subcluster, from west to east. The tail of the subcluster may be bent and blown eastward by this flow, which could have a radial gradient. In the same time, the ordered magnetic field lines, stretching from west-to-east could suppress conduction and thermodynamic instabilities, keeping the northern edge smooth.

VLA radio observations presented by \citet{schenck14} have revealed a bright radio galaxy between the S subcluster and the main cluster. This galaxy appears both in the X-ray and optical image, and a wide radio tail extends to the east from the location of the galaxy. The tail is oriented parallel to the northern edge of the subcluster, which is compatible with the west-to-east velocity field and supports the above scenario.

The interaction between the sloshing gas and the tail of the subcluster can explain the overall properties of the system consistently. Considering that galaxy clusters grow by accretion, this phenomenon is likely to be universal. Simulations of gas sloshing have so far mainly focused on binary mergers within relatively small radii. Future large scale sloshing simulations conducted out to large radii with magnetic field, and multiple merger simulations focusing on the gas dynamics will enable us to investigate the gas interaction quantitatively and will bring a deeper understanding as to how the gas evolves to a relaxed system.

\subsubsection{Gas sloshing in the core of the main cluster}\label{section_core}
\begin{figure*}
 \begin{minipage}{0.195\hsize}
  \includegraphics[width=1.35in]{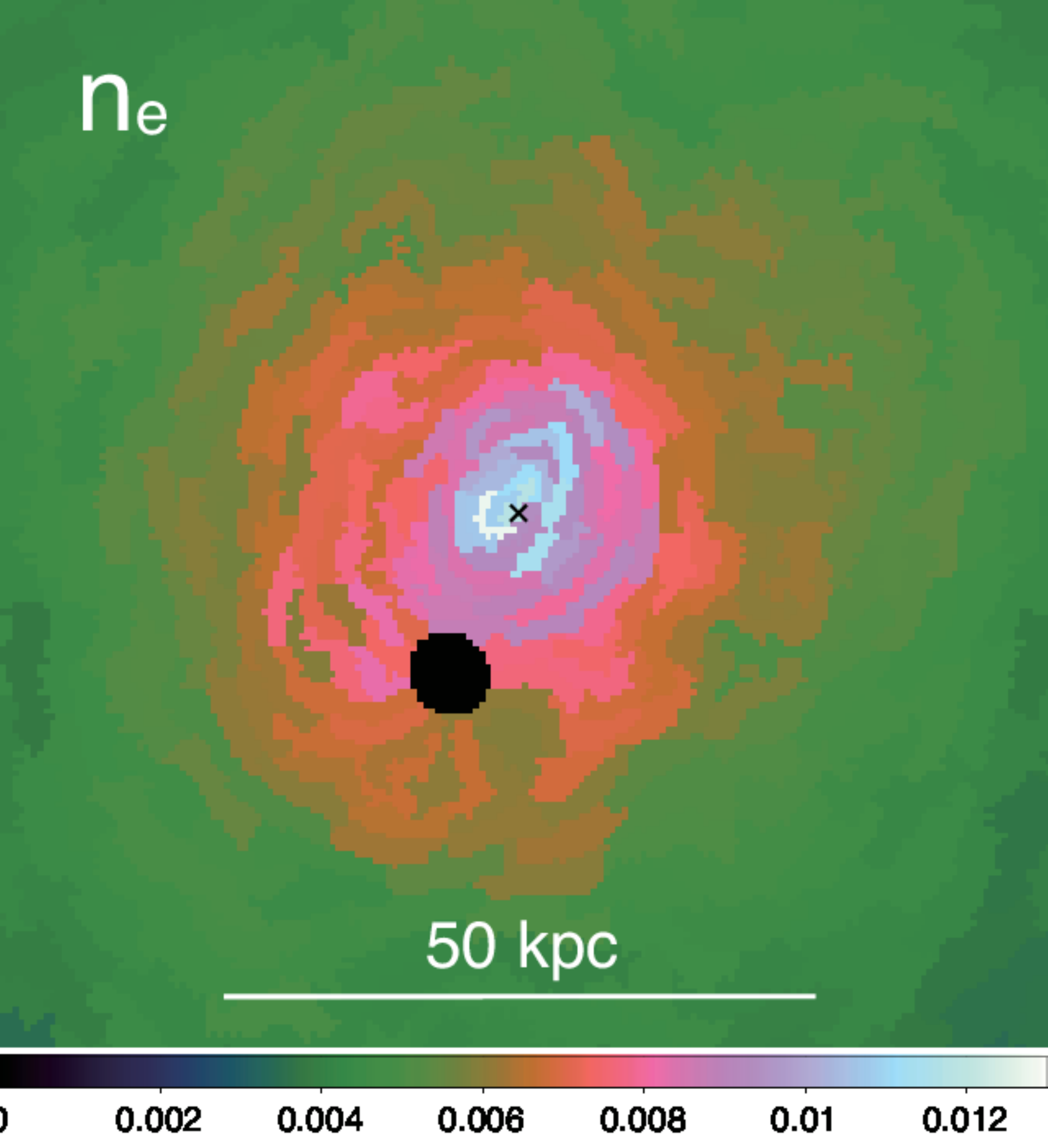}
 \end{minipage}
 \begin{minipage}{0.195\hsize}
  \includegraphics[width=1.35in]{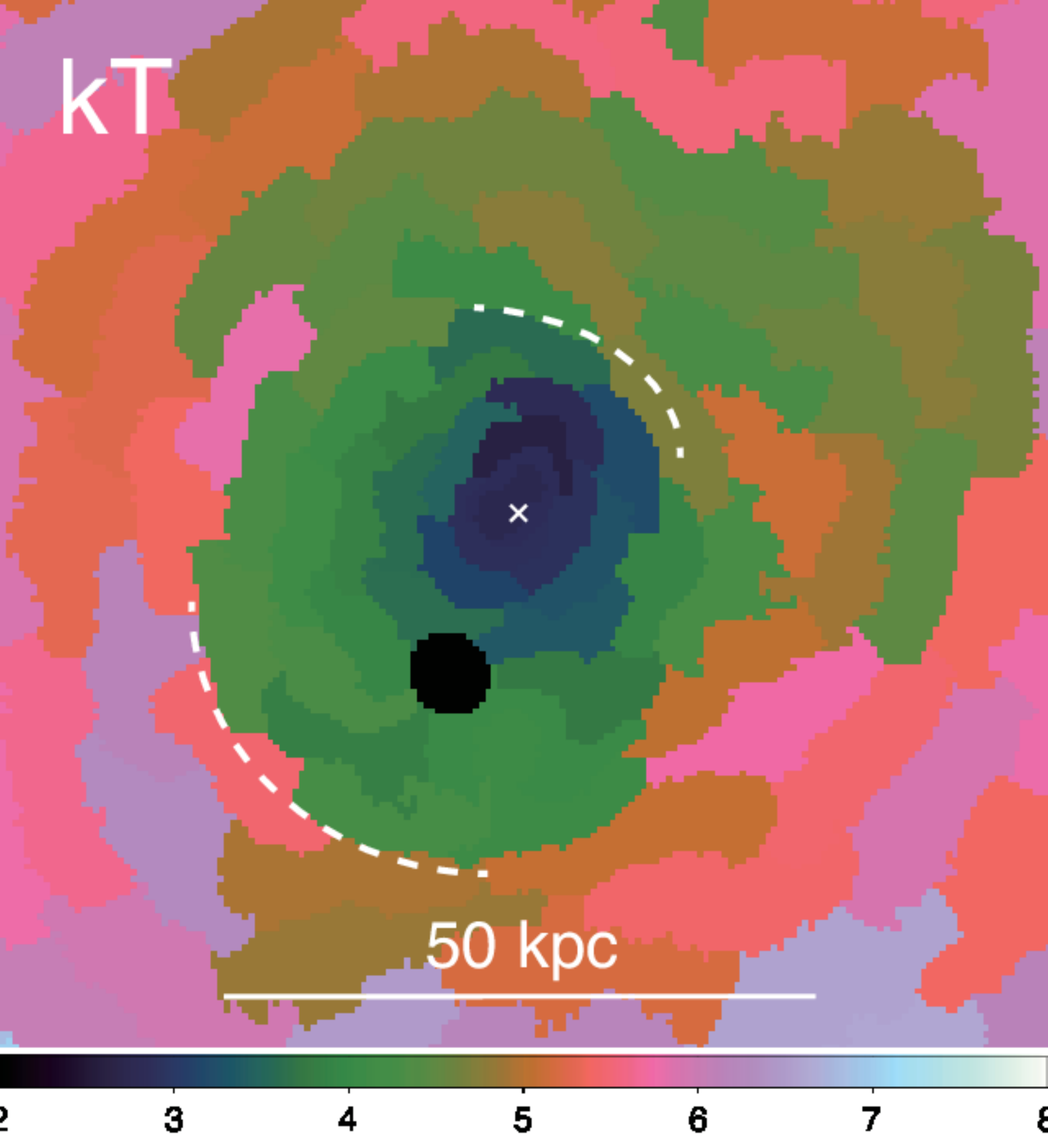}
 \end{minipage}
 \begin{minipage}{0.195\hsize}
  \includegraphics[width=1.35in]{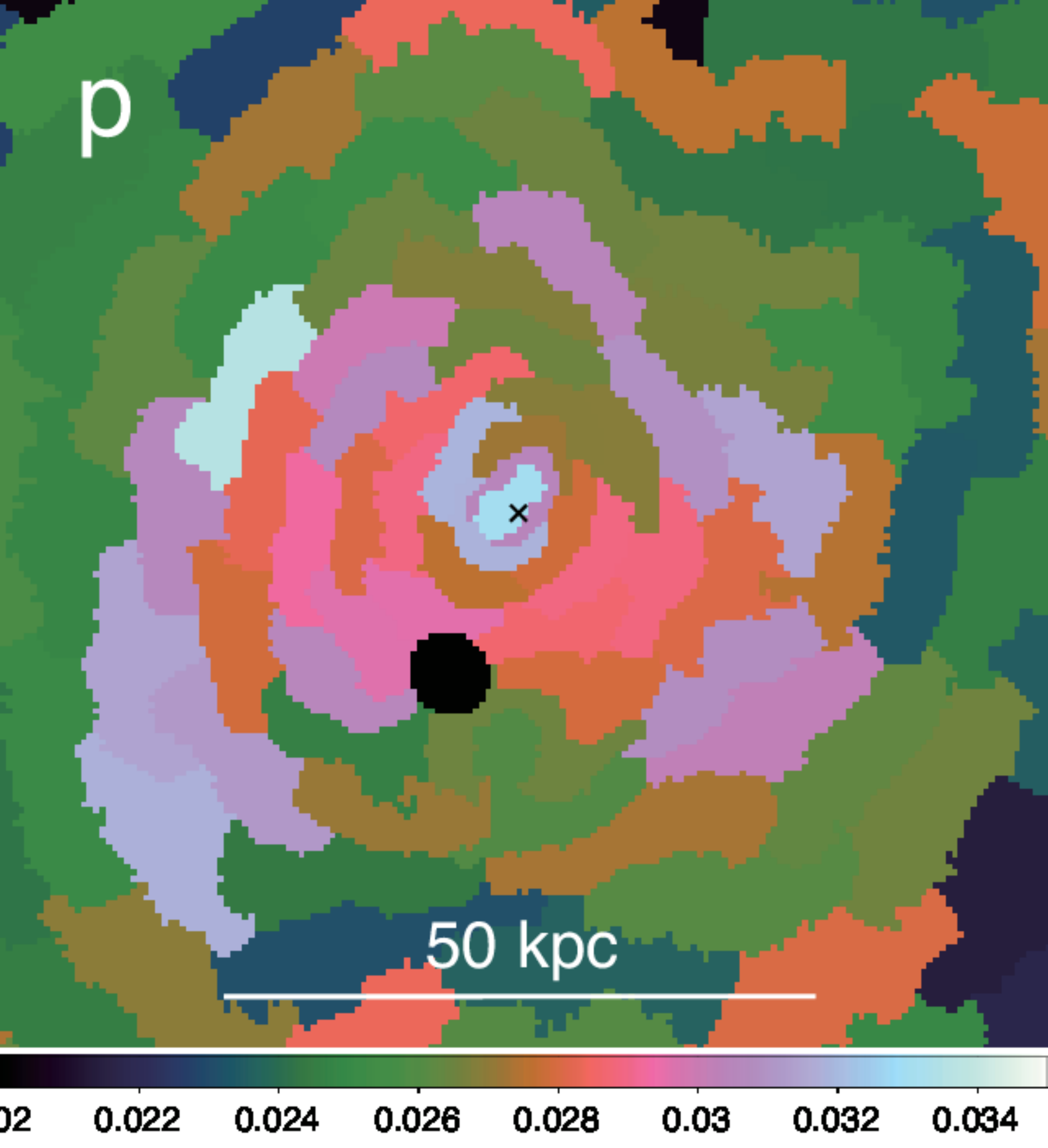}
 \end{minipage}
 \begin{minipage}{0.195\hsize}
  \includegraphics[width=1.35in]{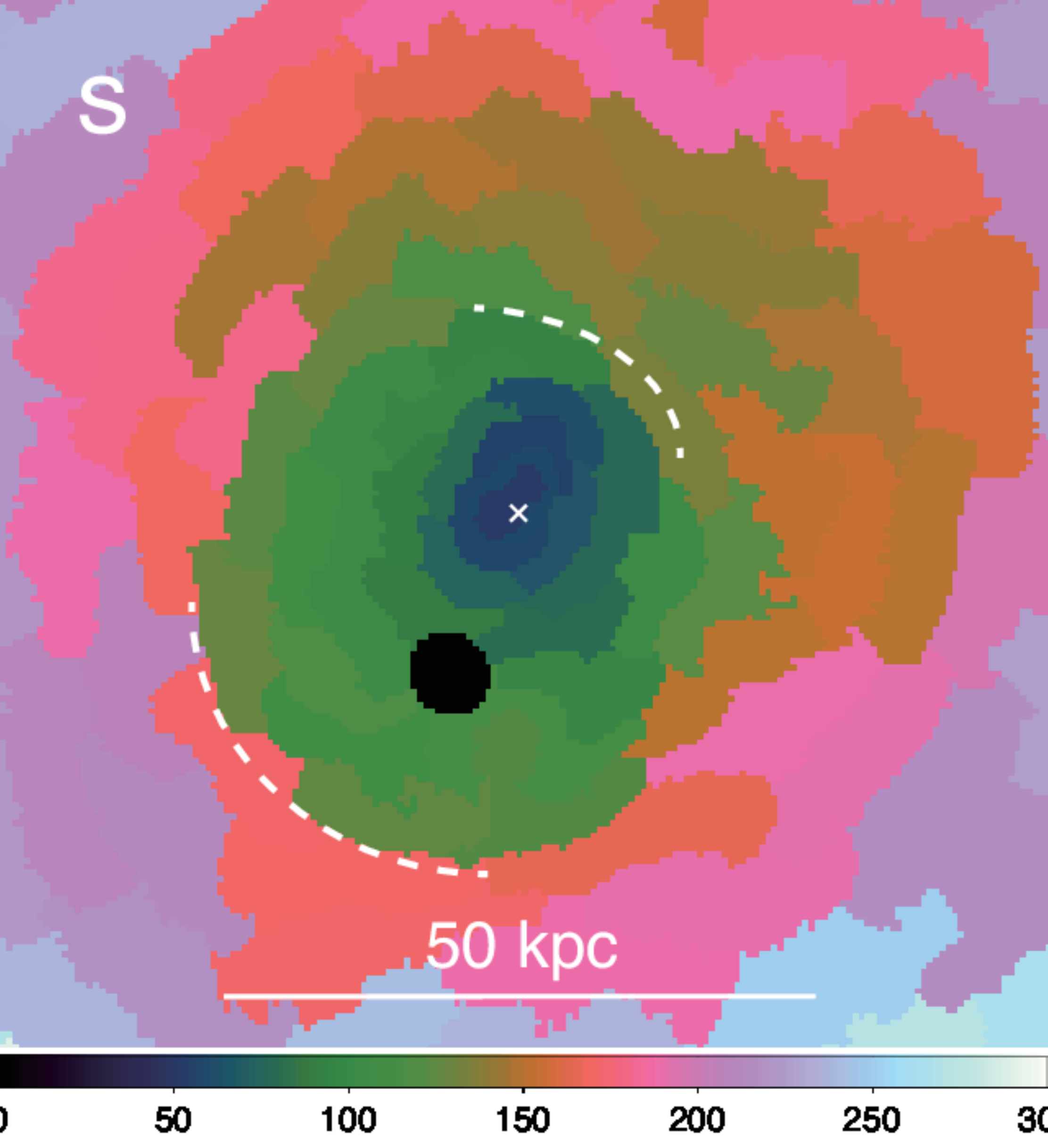}
 \end{minipage}
 \begin{minipage}{0.195\hsize}
  \includegraphics[width=1.35in]{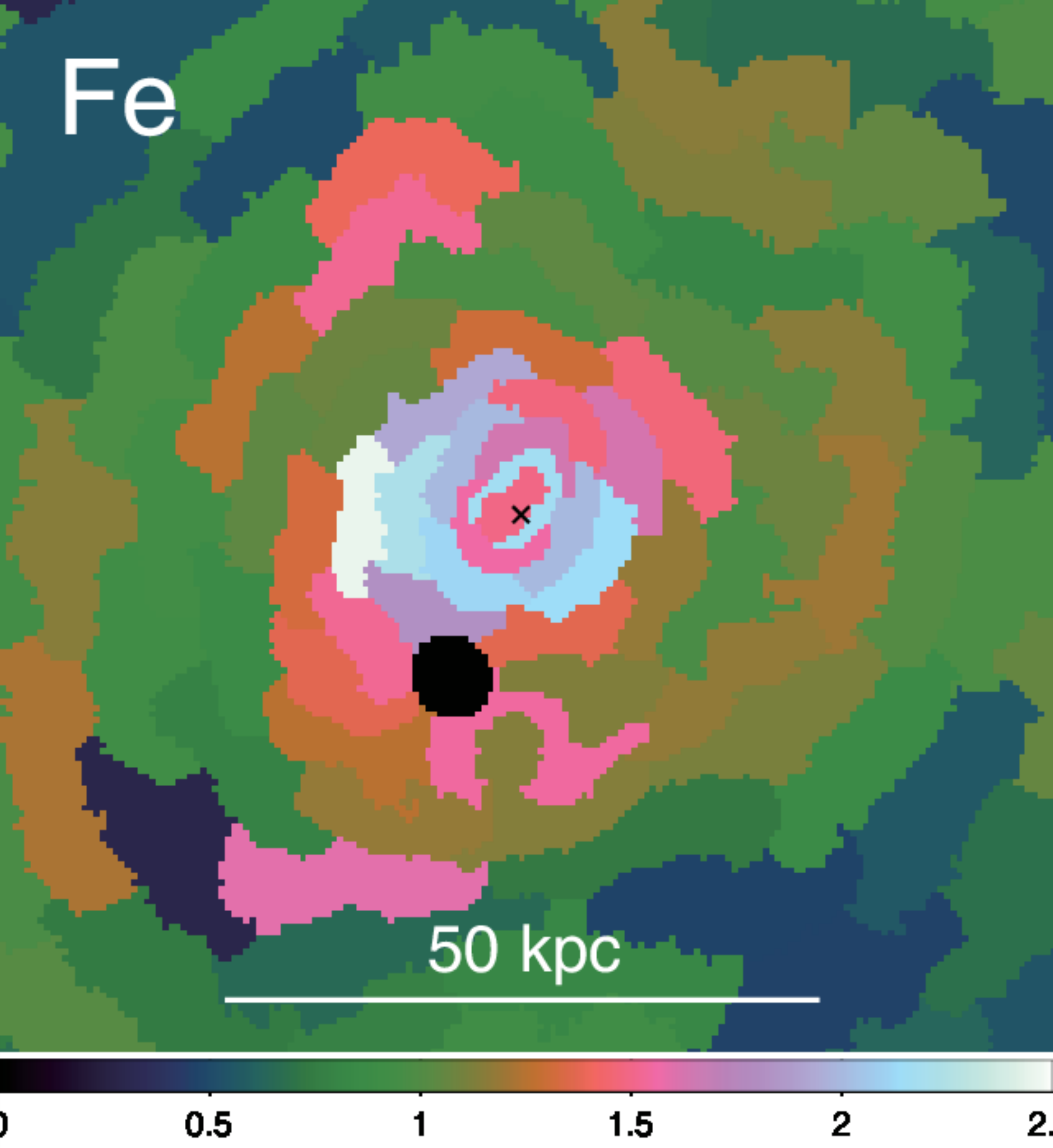}
 \end{minipage}
 \\
 \begin{minipage}{0.195\hsize}
  \includegraphics[width=1.35in]{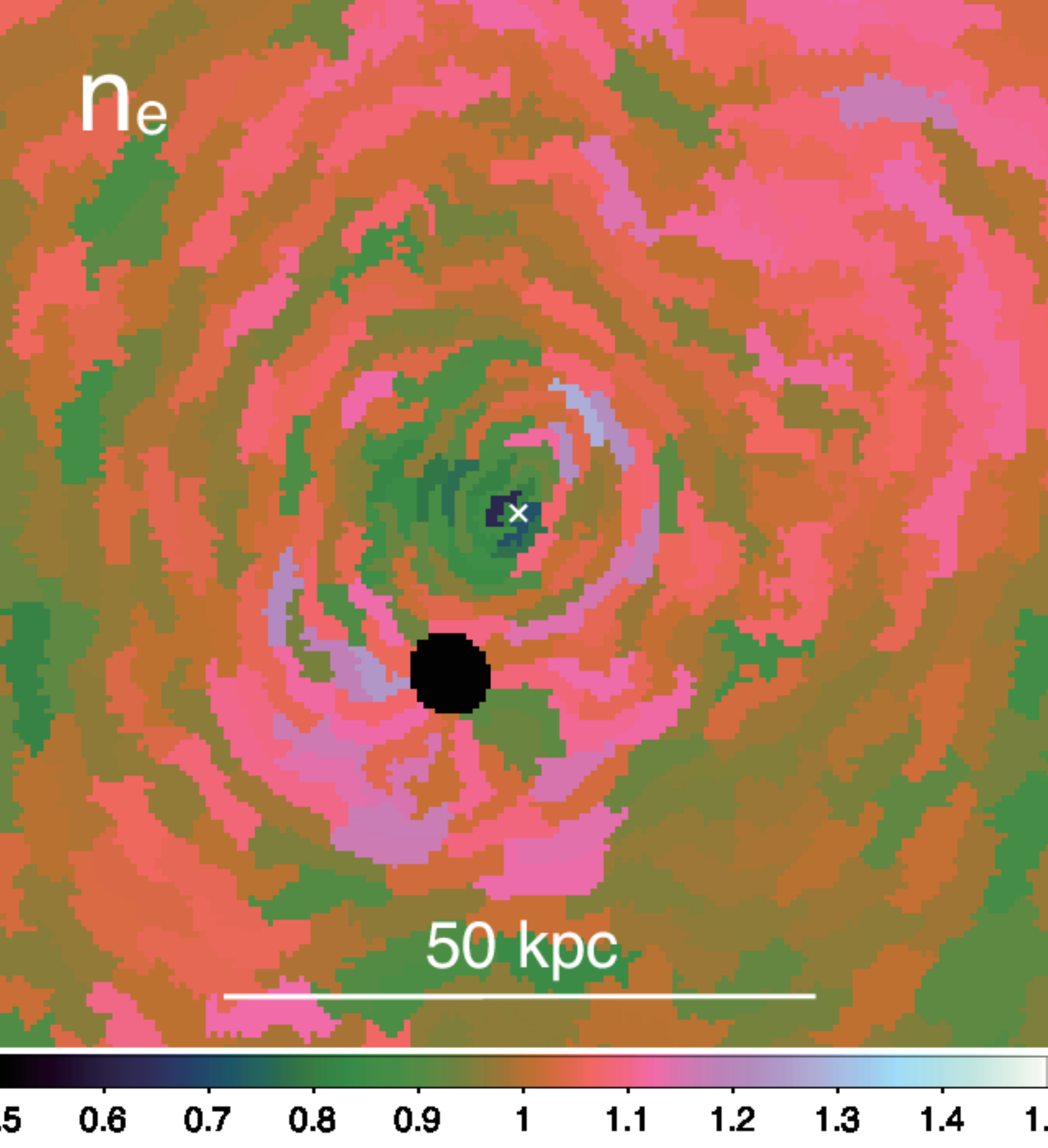}
 \end{minipage}
 \begin{minipage}{0.195\hsize}
  \includegraphics[width=1.35in]{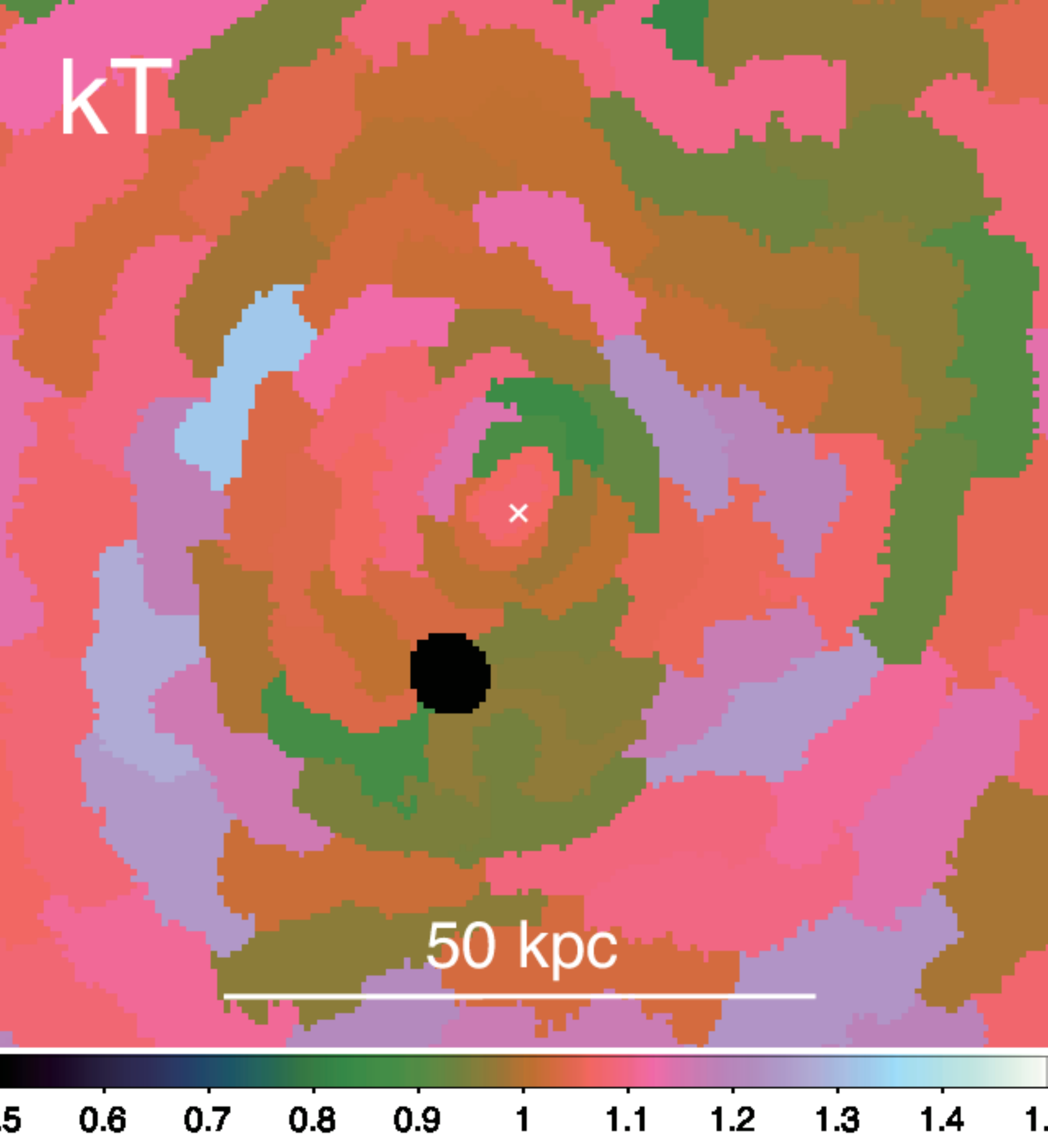}
 \end{minipage}
 \begin{minipage}{0.195\hsize}
  \includegraphics[width=1.35in]{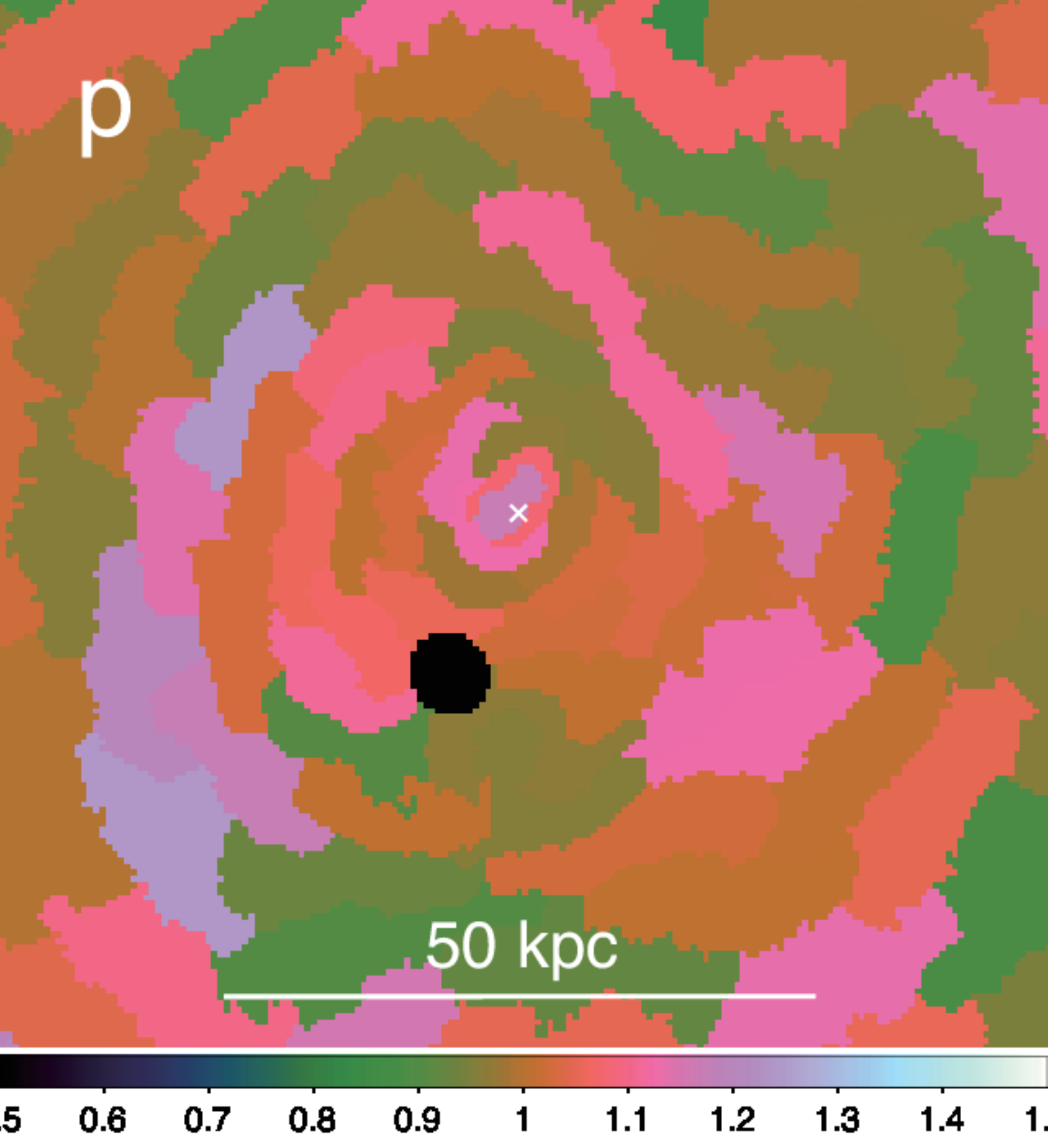}
 \end{minipage}
 \begin{minipage}{0.195\hsize}
  \includegraphics[width=1.35in]{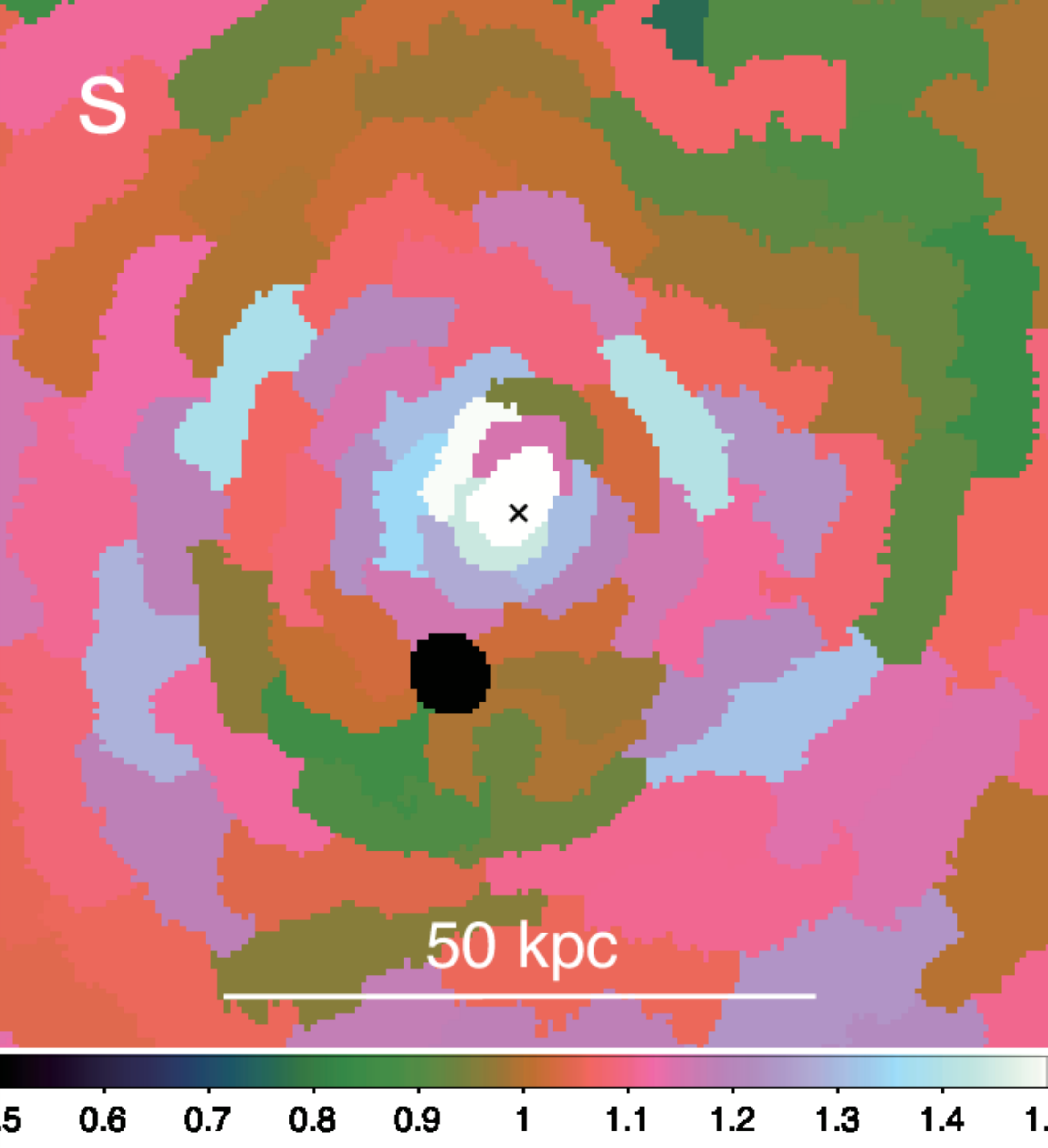}
 \end{minipage}
 \begin{minipage}{0.195\hsize}
  \includegraphics[width=1.35in]{figures/img_chandra_core_s4_2-ai.pdf}
 \end{minipage}
 \caption{Projected thermodynamic maps around the main cluster core. Upper five maps are the original maps and lower four are the trend-divided ones. Panels are the maps of density (cm$^{-3}\times(l/1~\mr{Mpc})^{-1/2}$), temperature (keV), pressure (keV cm$^{-3}\times(l/1~\mr{Mpc})^{-1/2})$ , entropy (keV cm$^2\times(l/1~\mr{Mpc})^{1/3})$ and Fe abundance in the unit of solar value respectively from left to right. The rightmost panel of the lower row is the closeup X-ray image of the core (see also Fig. \ref{img_core}). Typical errors are 5~per~cent for the density map, 20~per~cent for the Fe abundance map, and 10~per~cent for all other maps. The dashed curves represent the temperature and entropy edges.}
 \label{map_core}
\end{figure*}

The thermodynamic maps of the same region as Fig. \ref{img_core} are shown in Fig. \ref{map_core}. We see sharp edges at 15~kpc north and 35~kpc south of the main cluster core both in the temperature map and the entropy map, which correspond to the brightness edge in the image. They can be interpreted as cold fronts originating from the core gas sloshing, which is connected to the larger scale spiral out to several hundreds of kpc.

We detect a surface brightness cavity in the cluster core (Fig. \ref{img_core}), which is also known to host diffuse radio emission \citep{schenck14}. The surface brightness of the cavity is 0.87$\pm$0.02$\times10^{-6}$~ph~s$^{-1}$~cm$^{-2}$~arcsec$^{-2}$, which is significantly (more than 5$\sigma$) fainter than the surrounding ambient gas whose surface brightness is 1.03$\pm$0.02$\times10^{-6}$~ph~s$^{-1}$~cm$^{-2}$~arcsec$^{-2}$. The cavity is most likely due to a bubble blown by the AGN hosted by Holm 15A. The lack of a counterpart bubble may be due to the core gas sloshing which could hide the cavity.

\subsection{Shock features}\label{shock}
\subsubsection{The origin of the hotspot}\label{hotspot}

The only merging system that could have shock heated the gas to produce the ``hotspot'' is the infalling southern subcluster. Although the cool core remnant of this system is offset to the west of the hot spot, the dense low entropy gas associated with the tail seen in the Chandra image might be driving a shock as it is falling to the north.

If the hotspot (Fig. \ref{map}, Section \ref{ssub}) is due to previous shock heating, the required Mach number of the shock $\mathscr{M}$ can be calculated using the Rankine-Hugoniot jump condition. For $\gamma = 5/3$ gas, the ratio between the shock-heated gas temperature $T_2$ and the pre-shock gas temperature $T_1$ is determined through the equation
\begin{eqnarray}
 \dfrac{T_2}{T_1} &&= \dfrac{(\mathscr{M}^2+3)(5\mathscr{M}^2-1)}{16\mathscr{M^2}},
\end{eqnarray}
where $\mathscr{M}$ is the Mach number of the shock. In Fig. \ref{hotspot_regions}, we display the regions used to calculate the shock Mach number. The black wedge indicates the shock-heated region, and the white wedges are used to estimates the pre-shock gas temperature. The extracted temperatures are $kT_1 = 7.8\pm0.2$~keV and $kT_2 = 11.5^{+0.8}_{-0.7}$~keV, and the resulting Mach number is $\mathscr{M} = 1.5\pm0.1$, corresponding to a shock velocity of $\sim$2200~km~s$^{-1}$.

Since this pre-shock region is arbitrary, we evaluated the systematic uncertainty by splitting the pre-shock region into five smaller wedges as shown in Fig. \ref{hotspot_regions}. The obtained pre-shock temperature range of $kT_1 \approx 6$--9~keV, corresponds to a Mach number in the range of $\mathscr{M} \approx 1.3$--1.8, which is consistent with previous studies \citep[e.g.][]{tanaka10,schenck14}.

Although, currently the hotspot appears displaced from the line of the S subcluster's motion, it is most likely due to the shock driven by the infall of the S subcluster. Gas sloshing (Section \ref{sloshing}) can both redistribute the main cluster gas from an axisymmetric distribution and the associated ordered motion can displace the hotspot along the velocity field.

The mean line-of-sight velocity of Abell~85 is 16507 $\pm$ 102~km~s$^{-1}$ \citep{oegerle01}, and the line-of-sight velocity of the brightest central galaxy of the S subcluster is 16886 $\pm$ 35~km~s$^{-1}$ \citep{beers91}. The subcluster's relative line-of-sight velocity to the main cluster is thus estimated as 379 $\pm$ 107~km~s$^{-1}$. Assuming that the shock velocity of $\sim$2200~km~s$^{-1}$ represents the total velocity, the subcluster's motion is close to the plane of the sky.

\subsubsection{The dark band and the SW subcluster}\label{sw}
\begin{figure}
 \begin{center}
  \includegraphics[width=3.4in]{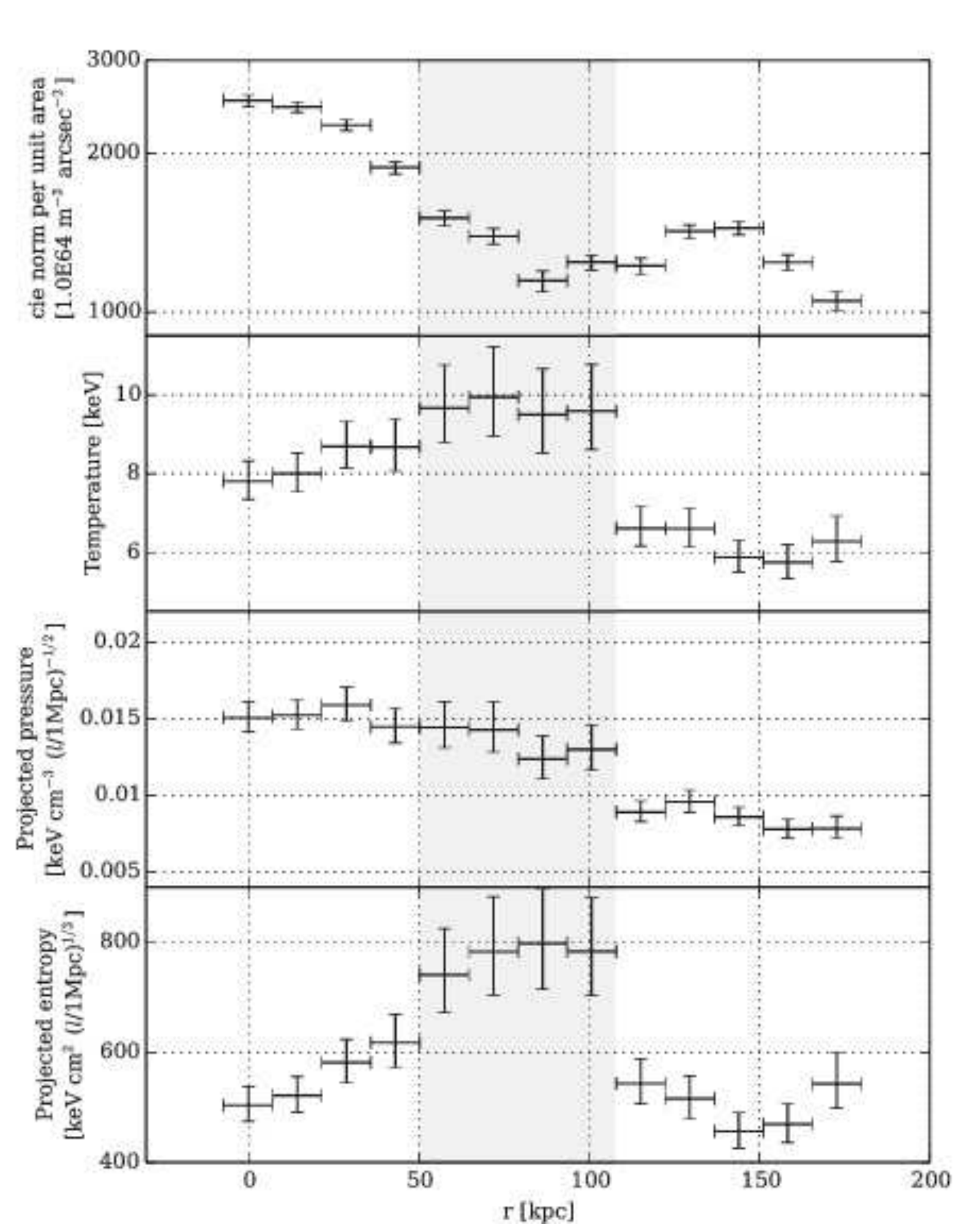}
 \end{center}
 \caption{Thermodynamic profiles along the line perpendicular to the interface between the sloshing arm and the SW subcluster. Regions are visualized in Fig. \ref{img_shock}. The grey rectangle represents the Dark band.}
 \label{shock_prof}
\end{figure}

The fact that the gas associated with the dark band is hotter than the gas at other azimuths at the same radius (approximately as hot as the southern hotspot) makes it likely that it has been shock heated. To investigate the profiles across the "dark band", we chose rectangular regions parallel to the interface (rectangles in Fig. \ref{img_shock}), and extracted the thermodynamic profiles shown in Fig. \ref{shock_prof}. In these profiles, the dark band is highlighted by a shaded grey rectangle.

We see clear excess both in the temperature and the entropy of the Dark band, which suggests the presence of shock-heated gas. However, because this region is the interface between the sloshing gas and the subcluster, it is difficult to see the thermodynamic contrast between the pre-shock gas and shock-heated gas clearly.

Assuming the gas has been shock heated, we estimate the Mach number $\mathscr{M}$ in the same way presented in the Section \ref{hotspot}. We combined the fifth, sixth, seventh and the eighth rectangular regions from left in Fig. \ref{img_shock} and \ref{shock_prof} as the shocked gas region and refitted the spectrum. The temperature derived for this region is $kT_2=9.1^{+0.5}_{-0.4}$~keV. As for the pre-shocked gas, we took a rectangular region which has the same shape and size as the shocked gas region, at the opposite position with respect to the central cD galaxy of the main cluster. The temperature of this region is $kT_1=6.5^{+0.4}_{-0.3}$~keV. The resulting Mach number is $\mathscr{M}=$1.4$\pm$0.1. 

\section{Conclusions}\label{conclusions}
With a deep 184~ks {\it Chandra} observation and complementary {\it XMM-Newton} and {\it Suzaku} data out to the virial radius, we studied the nearby, merging galaxy cluster system Abell~85 in an unprecedented detail. The main results of our work are summarized below:
\begin{itemize}
 \item We see a relatively large scale ($\sim$600~kpc) brightness excess spiral in the system, which strongly indicates that the ICM is sloshing in the gravitational potential of the cluster. This sloshing was likely triggered by previous merger events, in addition to the two currently ongoing mergers. 
 \item The S subcluster has a peculiar morphology with a clear southeastern tail that ends in an abrupt surface brightness drop. One of the edges of the tail is smooth for 200~kpc with the edge width of $\sim$10~kpc, which is significantly narrower than the Coulomb mean free path of the electrons in the main cluster gas, while the other edge is blurred and bent. 
 \item We propose a scenario which explains the overall properties of the system:
 \begin{enumerate}
  \item The sloshing was triggered several Gyrs ago, establishing ordered gas motions and ordered magnetic fields.
  \item The outer gas of the subcluster has been stripped earlier, and the low-entropy core gas is currently being stripped. 
   \item The stripped gas is being blown and bent by the ordered velocity field induced by the sloshing. At the same time, the ordered magnetic field lines that are stretched and oriented along the spiral, are preventing thermodynamic instabilities from developing across the sharp edge.
 \end{enumerate}
 \item The S subcluster core is almost entirely stripped of the low entropy gas, demonstrating a case of efficient destruction of a cool core during a merger.
 \item The tail of the S subcluster hosts X-ray bright gas clump candidates that are not associated with galaxies. The tail appears to continue out to around $r_{500}$ of the main cluster. The fact that the stripped tail of the infalling southern subcluster is seen across a radial range of over 700 kpc, indicates that the stripping of infalling subclusters may seed gas inhomogeneities.
 \item Beyond the stripped tail, which extends to $r_{500}$, the deprojected entropy profile along the infall direction is consistent with the theoretical prediction out to $r_{200}$. However, due to the large errorbars we cannot rule out flattening as observed in other systems.
 \item We confirm a previously known hotspot to the northeast of the S subcluster. The required Mach number to shock-heat the hotspot is in the range of $1.3 < \mathscr{M} < 1.8$, which is consistent with the previous observations.
 \item The interface between the SW subcluster and the main cluster has a high temperature and high entropy and might have been shock-heated by the merger of the subcluster. The estimated merger Mach number of the SW subcluster is $\mathscr{M} = 1.4\pm0.1$.
 \item The main cluster core hosts an AGN blown bubble/cavity in the south, and we see cold fronts around the core. The counterpart bubble may be hidden by the dense gas displaced by sloshing.
\end{itemize}

\section*{Acknowledgments}
Support for this work was provided by the National Aeronautics and Space Administration through Chandra Award Number GO3-14139X issued by the Chandra X-ray Observatory Center, which is operated by the Smithsonian Astrophysical Observatory for and on behalf of the National Aeronautics Space Administration under contract NAS8-03060. YI is financially supported by a Grant-in-Aid for Japan Society for the Promotion of Science (JSPS) Fellows. 
NW thanks the Japanese Institute of Space and Astronautical Science (ISAS) for hospitality. IY thanks the Kavli Institue for Particle Astrophysics and Cosmology (KIPAC) for hospitality. The authors thank the CHEERS collaboration for providing the deep XMM-Newton data of Abell 85. 
This work was supported in part by the U.S. Department of Energy under contract number DE-AC02-76SF00515.

\label{lastpage}

\end{document}